\documentclass[a4paper]{article}

\usepackage[dvips]{graphicx} 
\usepackage{color}           
\usepackage{amssymb}         
\usepackage{stmaryrd}        
\usepackage{varioref}        

\usepackage{mathlig}         
\input rules                 

\newcommand\ie{\textit{i.e.}}

\newcommand{\st}{\mbox{\textrm{s.t.}}}
\newcommand\newterm[1]{\emph{#1}}
\newcommand\COMMENT[1]{\mbox{\textit{\scriptsize#1}}}

\newcommand\bigstrut{}
\newcommand\PI[1]{\left(\bigstrut\Pi\,#1\right)}
\newcommand\SI[1]{\left(\bigstrut\Sigma\,#1\right)}
\newcommand\MU[1]{\left(\bigstrut\mu\,#1\right)}
\newcommand\NU[1]{\left(\bigstrut\nu\,#1\right)}
\newcommand\LAM[1]{\left(\bigstrut\lambda\,#1\right)}
\newcommand{\One}{\{\ast\}}
\newcommand\BinDown{\mathbin\downarrow}           
\newcommand\downclosure[1]{#1^{\downarrow}}       
\newcommand\upclosure[1]{#1^{\uparrow}}           

\newcommand\Set{\mathit{Set}}
\newcommand\Pow{\mathop\mathit{Pow}\nolimits}
\newcommand\Fam{\mathop\mathit{Fam}\nolimits}
\newcommand\BLANK{\_}
\newcommand\Sat[1]{\overline{#1}}

\newcommand\Id{\mathsf{id}}
\newcommand\eq{\mathsf{eq}}
\newcommand\Elim{\mathsf{Elim}}
\newcommand\Coiter{\mathsf{Coiter}}
\renewcommand\L{\mathrm{L}}
\newcommand\transition[1]{\stackrel{#1}{\longrightarrow}}

\newcommand{\Pos}{\mathbf{Pos}}                   
\newcommand{\gr}{\mathop{\mathbf{gr}}\nolimits}   

\newcommand\SET[2]{\ensuremath{\left\{\, #1\,\mid\, #2\,\right\}}}
\newcommand\FAM[2]{\SET{#1}{#2}}
\newcommand\SING[1]{\ensuremath{\left\{#1\right\}}} 
\newcommand\CHIP{\mathbin\setminus}
\newcommand\CHOP{\mathbin/}
\newcommand\SEQ{\fatsemi}

\newcommand\RULE[2]{\infer{#1}{#2}{}}

\newcommand\BFTop{\mathbf{BFTop}}
\newcommand\LinSim{\mathbf{LinSim}}
\newcommand\GenSim{\mathbf{GenSim}}
\newcommand\PT{\mathbf{PT}}

\newcommand{\TSI}[1]{\ensuremath{{#1}.A}}
\newcommand{\TSn}[1]{\ensuremath{{#1}.n}}
\newcommand{\ISA}[1]{\ensuremath{{#1}.A}}
\newcommand{\ISD}[1]{\ensuremath{{#1}.D}}
\newcommand{\ISn}[1]{\ensuremath{{#1}.n}}
\newcommand\RC{\ensuremath{\mathit{RC}}}
\newcommand\TC[1]{\ensuremath{{#1}^{+}}}
\newcommand\SKIP{\ensuremath{\mathsf{skip}}}
\newcommand\MAGIC{\ensuremath{\mathsf{magic}}}
\newcommand\ABORT{\ensuremath{\mathsf{abort}}}
\newcommand\angelU[1]{\ensuremath{\langle#1\rangle}}
\newcommand\demonU[1]{\ensuremath{[#1]}}
\newcommand\CONV[1]{\ensuremath{{#1}^{\sim}}}
\newcommand\RTC[1]{\ensuremath{#1^\ast}}
\newcommand\FISH[1]{\ensuremath{#1^\infty}}

\newcommand\restrict{\ensuremath\ltimes}      
\newcommand\cover{\ensuremath\mathrel{\lhd}}  
\newcommand\olp{\between}                     
\newcommand\A{\mathcal{A}}                    
\newcommand\J{\mathcal{J}}                    

\newcommand\sub{\subseteq}
\newcommand\comp{\complement}
\newcommand\sqsub{\sqsubseteq}
\newcommand\be{\[\begin{array}[t]{lllllll}}      \newcommand\ee{\end{array}\]}

\newcommand\INL[1]{\mbox{\ensuremath{\mathsf{in}}}_0(#1)}
\newcommand\INR[1]{\mbox{\ensuremath{\mathsf{in}}}_1(#1)}

\newcommand\constructor[1]{\mbox{\textsc{#1}}}
\newcommand\Cons{\constructor{cons}}
\newcommand\Call{\constructor{call}}
\newcommand\Exit{\constructor{exit}}
\newcommand\Nil{\constructor{nil}}
\newcommand\WHERE[1]{\ensuremath{\mathrel{\mbox{where}}\begin{array}[t]{l}#1\end{array}}}

\newcommand\exec{\ensuremath{\mathtt{exec}}}
\newcommand\Init{\ensuremath{\mathsf{Init}}}
\newcommand\Inv{\ensuremath{\mathsf{Inv}}}
\newcommand\Next{\ensuremath{\mathsf{Next}}}
\newcommand\Goal{\ensuremath{\mathsf{Goal}}}
 
\newcommand{\assert}[1]{\ensuremath{\angelU{#1}}}
\newcommand{\assume}[1]{\ensuremath{\demonU{#1}}}

\newcommand{\DATA}[1]{ \mbox{\rm\textbf{data}}\;\begin{array}[t]{ll}#1\end{array}}

\newenvironment{algproof}%
  {\begingroup\parindent=2cm\parskip=0pt plus 1pt
  \newcommand\step[2]{\leavevmode\llap{$##1$\enspace}\ifx\empty##2\else\hskip.25cm$\{$~{\scriptsize##2}~$\}$\fi}
  }
  {\endgroup}

\newtheorem{prop}{Proposition}
\newtheorem{lem}{Lemma}[section]
\newtheorem{cor}{Corollary}
\newtheorem{defn}{Definition}

\newcommand{\proofend}{\qed\endtrivlist}
\def\qed{\unskip\nobreak\hfil\penalty50\hskip1em\null\nobreak\hfil
  $\Box$\parfillskip=0pt\finalhyphendemerits=0\endgraf}
\newenvironment{proof}{\noindent\textbf{Proof:}}{\proofend}

\newenvironment{remark}%
  {\begin{list}{}{%
    \setlength{\topsep}{0pt}%
    \setlength{\leftmargin}{0pt}%
    \setlength{\rightmargin}{0pt}%
    \setlength{\listparindent}{\parindent}%
    \setlength{\itemindent}{\parindent}%
    \setlength{\parsep}{0pt plus 1pt}%
    \addtolength{\leftmargin}{1cm}%
    }\item%
    \footnotesize\medbreak\noindent\textbf{Remark.}\enspace}
  {\end{list}\medbreak}

\usepackage[leftbars]{changebar}
  {\par\noindent\bgroup\cbstart\small\textbf{TODO:}~\textsf\bgroup}%
  {\egroup\cbend\egroup\par}

\makeatletter
\def\timenow{\@tempcnta\time
  \@tempcntb\@tempcnta
    \divide\@tempcntb60
      \ifnum10>\@tempcntb0\fi\number\@tempcntb
        \multiply\@tempcntb60
          \advance\@tempcnta-\@tempcntb
            :\ifnum10>\@tempcnta0\fi\number\@tempcnta}
\makeatother

\title{Programming Interfaces and Basic Topology}

\author{Peter Hancock\protect\footnote{\texttt{hancock@spamcop.net}}
        \ and
        Pierre Hyvernat\protect\footnote{\texttt{hyvernat@iml.univ-mrs.fr}}}

\date{{\small\texttt{VERSION of \today\ at \timenow}}}

\begin{document}

  \mathlig{->}{\to}
  \mathlig{=>}{\Rightarrow}
  \mathlig{<}{\langle}
  \mathlig{>}{\rangle}
  \mathlig{<}{(}
  \mathlig{>}{)}
  \mathlig{<=>}{\Leftrightarrow}
  \mathlig{<==>}{\Longleftrightarrow}
  \mathlig{|-}{\vdash}
  \mathlig{-|}{\dashv}
  \mathlig{==}{\stackrel{\scriptscriptstyle \Delta}{=}}
  \mathlig{|=}{\ensuremath{\mathrel{\epsilon}}}
  \mathlig{×}{\times}
  \mathlig{|->}{\mapsto}
  \mathlig{<|}{\cover}
  \mathlig{|><}{\restrict}
  \mathlig{><}{\ensuremath{\times}}
  \mathlig{·}{\cdot}
  \mathlig{-o}{\mathbin{\relbar\mskip-8mu\circ}}

\maketitle

\begin{abstract}
  A pattern of interaction that arises again and again in programming, is a
  ``handshake'', in which two agents exchange data. The exchange is thought of
  as provision of a service.  Each interaction is initiated by a specific
  agent ---the client or Angel, and concluded by the other
  ---the server or Demon.

  We present a category in which the objects ---called interaction structures
  in the paper--- serve as descriptions of services provided across such
  handshaken interfaces.  The morphisms ---called (general) simulations---
  model components that provide one such service, relying on another.  The
  morphisms are relations between the underlying sets of the interaction
  structures. The proof that a relation is a simulation can serve (in
  principle) as an executable program, whose specification is that it provides
  the service described by its domain, given an implementation of the service
  described by its codomain.

  This category is then shown to coincide with the subcategory of ``generated''
  basic topologies in Sambin's terminology, where a basic topology is
  given by a closure operator whose induced sup-lattice structure need not be
  distributive; and moreover, this operator is inductively generated from a 
  basic cover relation. This coincidence provides topologists with a natural source of
  examples for non-distributive formal topology.  It raises a number of
  questions of interest both for formal topology and programming.

  The extra structure needed to make such a basic topology into a real
  \emph{formal topology} is then interpreted in the context of interaction
  structures.
\end{abstract}

\newpage
{\small \tableofcontents}
\newpage


\section{Introduction, preliminaries and notation} \label{sec:notation}

Programmers rarely write self-standing programs, but rather modules or
components in a complete system.  The boundaries of components are known as
interfaces, and these usually take the form of collections of procedures.
Commonly, a component exports or implements a ``high-level'' interface (for
example files and directory trees in a file system) by making use of another
``low-level'' interface (for example segments of magnetic media on disk
drives).  There is, as it were, a conditional guarantee: the exported
interface will work properly \emph{provided} that the imported one works
properly.

One picture for the programmer's task is therefore this:
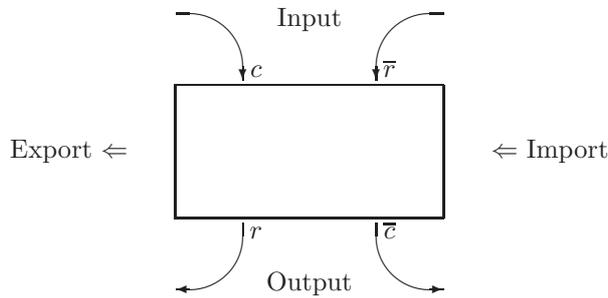
\begin{center}
\begin{picture}(345,130)(-25,80)
  \put(100,125){\framebox(100,50){}}
  \put(60,150){\makebox(0,0){\shortstack{Export $\Leftarrow$}}}
  \put(240,150){\makebox(0,0){\shortstack{$\Leftarrow$ Import}}}
  \put(100,177){\oval(50,50)[tr]}
  \put(125,177){\vector(0,-1){0}}
  \put(125,178){\makebox(0,0)[b]{$\quad c$}}
  \put(200,177){\oval(50,50)[tl]}
  \put(175,177){\vector(0,-1){0}}
  \put(175,178){\makebox(0,0)[b]{$\quad \overline{r}$}}
  \put(200,123){\oval(50,50)[bl]}
  \put(200,98){\vector(1,0){0}}
  \put(175,117){\makebox(0,0)[b]{$\quad \overline{c}$}}
  \put(100,123){\oval(50,50)[br]}
  \put(100,98){\vector(-1,0){0}}
  \put(125,117){\makebox(0,0)[b]{$\quad r$}}
  \put(150,200){\makebox(0,0){Input}}
  \put(150,100){\makebox(0,0){Output}}
\end{picture}
\end{center}

The task is to ``fill the box''.  In this picture the horizontal dimension
shows interfaces.  The exported, higher-level interface is at the left
and the imported, lower-level interface at the right.
The vertical dimension shows communication events (calls to and returns from
procedures), with data flowing from top to bottom: $c$ and $\overline{r}$
communicate data from the environment, while $r$ and $\overline{c}$
communicate data to the environment.  The labels $c$ (for command or call) and
$r$ (for response or return) constitute events in the higher level interface,
while $\overline{c}$ and $\overline{r}$ are at the lower level.
The pattern of communication is that first there is a call to the command~$c$,
then some number of repetitions of interaction pairs
$\overline{c}\overline{r}$, then finally a return $r$.

The picture this gives of the assembly of a complete system is that one has a
series of boxes, with input arrows linked to output arrows by a ``twisted pair
of wires'' reminiscent of the Greek letter ``$\chi$''\label{chi}.  This is indeed a kind
of composition in the categorical sense, where the morphisms are components.
The paper is about this category.

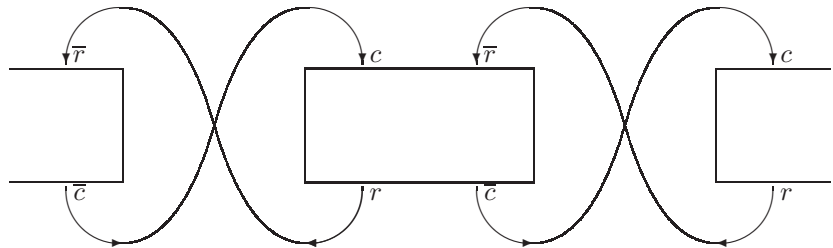
\begin{center}
\begin{picture}(345,100)(-50,75)
  \setlength{\unitlength}{0.30mm}
  \put(100,125){\framebox(100,50){}}
  \put(-5,178){\makebox(0,0)[b]{$\quad \overline{r}$}}
  \put(-5,117){\makebox(0,0)[b]{$\quad \overline{c}$}}
  \put(20,177){\oval(50,50)[tl]}
  \put(20,123){\oval(50,50)[bl]}
  \put(-5,177){\vector(0,-1){0}}
  \put(20,175){\line(-1,0){50}}
  \put(20,125){\line(0,1){50}}
  \put(20,125){\line(-1,0){50}}
  \put(20,98){\vector(1,0){0}}
  \qbezier(20,202)(45,202)(60,150)
  \qbezier(60,150)(75,98)(100,98)
  \put(100,123){\oval(50,50)[br]}
  \put(100,98){\vector(-1,0){0}}
  \qbezier(20,98)(45,98)(60,150)
  \qbezier(60,150)(75,202)(100,202)
  \put(100,177){\oval(50,50)[tr]}
  \put(125,177){\vector(0,-1){0}}
  \put(125,178){\makebox(0,0)[b]{$\quad c$}}
  \put(200,177){\oval(50,50)[tl]}
  \put(175,177){\vector(0,-1){0}}
  \put(175,178){\makebox(0,0)[b]{$\quad \overline{r}$}}
  \put(200,123){\oval(50,50)[bl]}
  \put(200,98){\vector(1,0){0}}
  \put(175,117){\makebox(0,0)[b]{$\quad \overline{c}$}}
  \put(100,123){\oval(50,50)[br]}
  \put(100,98){\vector(-1,0){0}}
  \put(125,117){\makebox(0,0)[b]{$\quad r$}}
  \qbezier(200,202)(225,202)(240,150)
  \qbezier(240,150)(255,98)(280,98)
  \put(280,123){\oval(50,50)[br]}
  \put(280,98){\vector(-1,0){0}}
  \qbezier(200,98)(225,98)(240,150)
  \qbezier(240,150)(255,202)(280,202)
  \put(280,177){\oval(50,50)[tr]}
  \put(305,177){\vector(0,-1){0}}
  \put(280,175){\line(1,0){50}}
  \put(280,125){\line(0,1){50}}
  \put(280,125){\line(1,0){50}}
  \put(305,178){\makebox(0,0)[b]{$\quad c$}}
  \put(305,117){\makebox(0,0)[b]{$\quad r$}}
\end{picture}
\end{center}

How can we describe interfaces?  Interface description languages (such as IDL
from \texttt{http://www.omg.org/}) commonly take the form of signatures, \ie~typed procedure
declarations.  The type system is ``simply typed'', and it is used in
connection with encoding and decoding arguments for possible remote
transmission.  It addresses other mechanistic, low-level and administrative
issues.  However an interface description ought to describe everything
necessary to design and verify the correctness of a program that uses the
interface, without knowing anything about how it might be implemented.  It
should state with complete precision a \emph{contract}, or in Dijkstra's words
a ``logical firewall'' between the user and implementer of an interface.

We define this category in (essentially) Martin-L\"of's type theory, a
constructive and predicative type theory in which the type-structure, is
sufficiently rich to express specifications of interfaces with full precision.
One reason for working in a constructive type theory is that a model for
program components in such a setting is \textit{ipso facto} a ``working''
model.  In principle, one may write executable program components in this
framework, and exploit type-checking to ensure that they behave correctly.  In
practice, one has to code programs in real programming languages.
Nevertheless, one can perhaps develop programs in a dependently typed
framework, using type-checking to guide and assist the development (as it were a
mental prosthesis), run the programs to debug the specifications, 
and then code the programs in a real programming notation.


Our model is constructed from well-known ingredients.  Since the seminal work
of Floyd, Dijkstra and Hoare \cite{Floyd67, dijkstra76:_discip_progr, Hoare69}
there has been a well established tradition of specifying commands in
programming languages through use of predicate transformers, and roughly
speaking the objects of our category are predicate transformers on a
state-space.  Equally well established is the use of simulation relations to
verify implementations of abstract data types, and roughly speaking, the
morphisms of our category are simulation relations, or more precisely,
relations together with a proof that they are simulations.  The computational
content of a simulation is contained in this (constructive) proof.

However, the ``natural habitat'' of the notions of predicate transformer and
simulation is higher-order (impredicative) logic.  To express these notions in
a \emph{predicative} framework, we work instead with concrete, first-order
representations in which their computational content is made fully explicit.
Again, the key ideas are fairly well-known, this time in the literature of
constructive mathematics and predicative type theory.  Our contribution is
only to put them to use in connection with imperative programming.

Finally, our excuse for submitting a paper on programming to a conference on
formal topology is that our category of interfaces and components turns out to
coincide almost exactly with the category of basic topologies and basic
continuous relations in Sambin's approach to formal topology.  At the least,
one can hope that further development of this approach to program development
can benefit from research in the field of formal topology.  One may also hope
that work in formal topology can benefit in some way  from several decades of
intensive research in the foundations of imperative programming and perhaps
even gain a new application area.

\subsection{Plan of the paper}


The first main section (\ref{sec:subsets-relations}) begins with 
two ways in which the notion of subset can be expressed in type theory.  Then
set up some machinery for dealing with binary relations, to
illustrate how our notions of subset have repercussions on higher order
notions.  In essence, we obtain besides the ordinary notion of relation a more
computationally oriented notion of \emph{transition structure}, that
pre-figures our representation of predicate transformers.

The next two sections (\ref{sec:PT} and \ref{sec:IS}) concern the notion of
monotone predicate transformer. 
In the first of
these sections (\ref{sec:PT}), we review the notion of predicate transformer
as it occurs in the theory of inductive definitions and in the semantics of
imperative programming.  The main points here are that predicate transformers
form a complete lattice under pointwise inclusion, that they possess also a
monoidal structure of sequential composition, and moreover that there are two
natural forms of ``iteration''.
Section~\ref{sec:IS} is devoted to a predicative analysis of the notion of
predicate transformer.  This exploits the distinction drawn in section
\ref{sec:subsets-relations} between our two forms of the notion of subset.  We
represent predicate transformers by objects called \emph{interaction
structures,} and show that our representations supports the same algebraic
structure.

The objects of our category are interaction structures over a set of states.
The next section (section \ref{sec:morphisms}) is about morphisms between
these objects.  It is convenient to unfold our answer in three stages.  In the
first step we define a restricted notion of \emph{linear simulation} (that is
indeed connected with the linear implication of linear logic) for which an
interaction in the domain is simulated by exactly one interaction in the
codomain.  In the second step, we move to the Kleisli category for a monad
connected with the reflexive and transitive closure of an interaction
structure; we call the morphisms in the Kleisli category \emph{general
simulations.}  In the third and last step, taking a hint from formal topology,
we take a quotient of general simulations, by passing to the \emph{saturation}
of a relation.  The last step captures the idea that two relations may have
the same simulating potential, modulo some hidden interactions.

Up to this point, the constructions have been motivated by
considerations from imperative programming.  In section
\ref{sec:topology}, we examine the connection with formal topology.
Firstly, our category of interaction structures and general morphisms
corresponds exactly to Sambin's category of inductively defined basic
topologies.  Secondly, formal topology goes beyond basic topology by
adding a notion of convergence, that allows for an analysis of the
notion of point.  The remainder of section \ref{sec:topology} is
concerned with a tentative interpretation of this extra structure.

We conclude with some questions raised in the course of the paper, and
acknowledgment of some of the main sources of our ideas.

\subsection{Mathematical framework(s)}

We work in a number of different foundational settings, that we have tried to
stratify in the following list.

\begin{itemize}

\item At the bottom, the most austere is Martin-L\"of's type theory
  (\cite{ML84,NPS}), with a principle of inductive definitions similar
  to that used by Petersson and Synek in the paper \cite{PS89}, with certain
  forms of universe type, but without any form of propositional equality.

  Our category of interfaces and components can be defined using only
  predicative type theory with inductive definitions.
  In fact the category has been defined and its basic properties proved in
  such a theory using the Agda ``programming'' language (\cite{Agda}). The
  proof scripts can be found at
  \texttt{http://iml.univ-mrs.fr/~hyvernat/academics.html}.

\item To this we add rules for propositional equality, which is
  necessary to round-out the programming environment to a language for fully
  constructive (intuitionistic and predicative) mathematics.

  This is not the right place to try to analyze the notion of equality, in any
  of its manifestations: definitional, propositional, judgmental, intensional,
  extensional and so on.  It is however a source of non-computational
  phenomena in type theory, and the history of predicative type theory (if not
  also its future) is one of a constant struggle with this notion.  We wish to
  carefully track the use of the equality relation (and cognate notions such
  as singleton predicate).  That is we prefer to work with ``pre-sets'' rather
  than ``setoids'' \cite{Hofmann94}.

  \item We also add a principle for coinductive definitions.
  The foundations of coinduction in predicative mathematics are not yet
  entirely clear.  We simply use co-inductive definitions in the most
  ``straightforward'' way, meaning by this that our constructs seem to make
  good computational sense.  One reference for the kind of coinductive
  definitions we will use 
  can be found in~\cite{setzerhancock:venice2003}.

\item At various points, it seems necessary to relax the stricture of
  predicativity.  In particular, we invoke the Knaster-Tarski theorem. This
  lacks a strictly predicative justification.  Since we are trying to devise
  computationally-oriented analogues of certain impredicative constructions,
  it is necessary to look at matters from the impredicative point of view, if
  only for comparison.

\item Finally, at the highest or most abstruse level, we shall
  occasionally make use of classical, impredicative reasoning, thus going
  beyond any straightforward computational interpretation.  Working at this
  level Hyvernat (\cite{PTlinear,PTlinearSecondOrder}) has identified
  surprising connections between an impredicative variant of our category and
  classical linear logic, even of second order.
\end{itemize}

\subsection{Type theoretic notation}

Our notation is based (loosely) on Martin-L\"of's type theory, as expounded
for example in \cite{ML84,NPS}.  In the paper we call this simply ``type
theory''.

\begin{itemize}

\item To say that a value $v$ is an element of a set $S$, we write $v
  \in S$.  On the other hand, to say that $o$ is an object of a proper type
  $T$ (such as $\Set$, the type of sets), we write $o : T$.

\item We use standard notation as in, for example \cite{ML84,NPS},
  for indexed cartesian products and disjoint unions. This is summarized in
  the following table:

\smallbreak
  \begin{center}
  \begin{tabular}{l|l|l}
                           & product             &  sum                    \\
    \hline
    dependent version      & $\PI{a\in A}{B(a)}$ & $\SI{a\in A}{B(a)}$     \\
    non-dependent          & $A -> B$            & $A×B$                   \\
    element in normal form & $\LAM{a\in A}b$     & $<a,b>$                 \\
  \end{tabular}
  \end{center}
  We iterate those constructions with a comma.  Using the Curry-Howard
  isomorphism, we might also use the logical $\forall$ and $\exists$ as
  notations for $\Pi$ and~$\Sigma$.
  \\
  We use the same notation at the type level.

\item Instead of the binary disjoint union $A + B$, we prefer to use a
  notation in which constructors can be given mnemonic names, as is common in
  programming environments based on type theory. For example, the disjoint
  union $A+B$ itself could be written $\mbox{\rm\textbf{data}}\;\INL{a \in A}
  \ |\  \INR{b \in B}$.  As the eliminative counterpart of this construction,
  we use pattern matching.

    We also use ad-lib pattern matching in defining functions by recursion,
    rather than explicit elimination rules (recursors, or ``weakly initial
    arrows'').

  \item We use simultaneous inductive definitions of a family of sets over a
    fixed index-set (as in~\cite{PS89,NPS}), with similar conventions.

    At an impredicative level, we will make use of $\mu$-expressions for
    inductively defined sets, predicates, relations, and predicate
    transformers.



\end{itemize}

\section{Two notions of subset}
\label{sec:subsets-relations}

We will be concerned with two notions of subset, or more accurately two forms
in which a subset of a set $S$ may be given:
\be
  \SET{ s \in S }{ U(s) } \qquad \hbox{or}\qquad
  \FAM{ f(i) }{ i \in I }\ \hbox{.}
\ee
The first we call ``predicate form'' ---$U$ is a predicate or propositional
function with domain $S$.  The second we call ``indexed form'', or ``family
form'' ---$f$ is a function from the index set $I$ into $S$.  Other
terminology might be ``comprehension'' versus ``parametric'', or
``characteristic'' versus ``exhaustive''.

For example, here are two ways to give the unit circle in the Euclidean plane:
(note that we do not require in indexed form that the function $f$ is
injective)
\[
  \SET{ (x,y) \in \mathbb{R}^2 }{ x^2 + y^2 = 1 } \qquad \hbox{or}\qquad
  \FAM{ (\sin \theta, \cos \theta) }{ \theta \in \mathbb{R} } \ \hbox{.}
\]
Of course, what we write in one form we may write in the other:
\be
  \FAM{ s }{ <s,\BLANK> \in \SI{ s \in S} U(s) } \ \hbox{;}   &\quad  \COMMENT{(predicate rewritten as family)} \\
  \SET{ s \in S }{ (\exists i \in I)\, s =_S f(i) } \ \hbox{.} &\quad \COMMENT{(family rewritten as predicate)}
\ee
To turn a predicate into an indexed family, we take as index the set of
proofs that some elements satisfy the predicate, and for the indexing function
the first projection.
To turn an indexed family into a predicate, we make use of the equality
relation ``$=_S$'' between elements of $S$, and in essence form the union of a
family of singleton predicates: $\bigcup_{i \in I}\SING{f(i)}$.

\smallbreak
So it may seem that what we have here is a distinction without any real
difference.  Note however that the essence of a predicate is a (set-valued)
function defined \emph{on}~$S$, while the essence of an indexed family is a
function \emph{into}~$S$, so that there is a difference in variance.  To make
this clear, let us define two functors which take a set~$S$ to the type of
predicate-form subsets of~$S$, and to the type of indexed-form subsets of~$S$.

\begin{defn} Define the following operations:
\begin{itemize}
\item
  $\Pow(S) == S -> \Set$, where we may write $U:\Pow(S)$ as $\SET{s\in S}{U(s)}$;
  and if $f\in S_1 -> S_2$, then\\
  $\begin{array}[t]{lllll}
  \Pow(f) & : & \Pow(S_2) & ->  & \Pow(S_1)  \\
          &   & U         & |-> & \SET{ s_1 \in S_1 }{ U\big(f(s_1)\big) }
   \end{array}$
  \smallbreak
  We write $U\sub S$ as a synonym for $U:\Pow(S)$. Note that $\Pow(f)$ is
  usually written $f^{-1}$.
\item
  $\Fam(S) == \SI{ I : \Set }S -> I$, where we may write  $<I,x>:\Fam(S)$ as $\FAM{x(i)}{i\in I}$;
  and if $f\in S_1 -> S_2$, then\\
  $\begin{array}[t]{lllll}
  \Fam(f) & : & \Fam(S_1)               & ->  & \Fam(S_2) \\
          &   & \FAM{ x(i) }{ i \in I } & |-> & \FAM{ f(x(i)) }{ i \in I }
  \end{array}$

\end{itemize}
\end{defn}
The first functor is contravariant, while the second is covariant.  So the
distinction we have made corresponds after all to a well-known (even banal)
difference.

In a predicative framework, both these functors cross a ``size'' boundary:
they go from the category of (small) sets to the category of (proper) types.
In fact these functors can be extended to endo-functors at the level of types,
going from the category of (proper) types to itself.  Remark however that the
translations between subsets and families can not be carried out in either
direction at the level of types.
\begin{itemize}

  \item Going from families to subsets would require a propositional
    (\ie~set-valued) equality relation between the objects of arbitrary types,
    rather than merely between the elements of a set.

  \item Going from a propositional function defined on a type to an indexed
    family is in general impossible since we require the indexing set to be
    \ldots\ a set.

\end{itemize}
This will become important when we iterate or compose our two variants of
the power-functor.

\medbreak
If we call into question, or try to work without the idea of a generic notion
of propositional equality, the two notions of subset fall into sharp relief.
In basic terms, the intuition of the distinction is that a family is something
computational, connected with what we ``do'' or produce.  On the other hand, a
predicate is something specificational, connected with what we ``say'' or
require.

\medbreak
How does the algebraic structure of predicates compare with that of indexed
families?  As for predicates, the situation is the normal one:  if we
interpret the logical constants constructively, they form a Heyting algebra.
With the equality relation, the lattice is atomic, with singleton predicates
for atoms.  The inclusion and ``overlap'' relations are defined as follows:

\begin{defn}
  Let $U$ and $V$ be two subsets of the same set $S$; define:
  \begin{itemize}
    \item $s |= U == U(s)$ (\ie~$s|=U$ iff ``\/$U(s)$ is inhabited'');
    \item $ U \sub V == \PI{s \in S} U(s) -> V(s)$;
      \hskip1cm\COMMENT{(\ie~$(\forall s\in S)\ s|=U -> s|=V$)}
    \item $ U \olp V == \SI{s \in S} U(s) \land V(s)$.
  \end{itemize}
\end{defn}
The importance of $\olp$ in a constructive setting has been stressed by
Sambin: it is a positive version of non-disjointness, dual to inclusion.

\begin{remark}
  The confusion between the two meanings of ``$\sub$'' can always be resolved
  (``$\sub$'' is a synonym for $\BLANK:\Pow(\BLANK)$ and denotes inclusion of
  subsets).  For a full account of traditional set theoretic notions in
  ``subset theory'', we refer to~\cite{toolbox}. Here are two examples:
  \begin{itemize}
    \item $S_{\mathrm{Full}} == \SET{s\in S}{\top}$ contains all the elements
      of $S$. We write it simply $S$;
    \item $U×V == \SET{(s,s')\in S×S}{s|=U\ \hbox{and}\ s'|=V}$.
  \end{itemize}
\end{remark}

\smallbreak
What now about families?  In the presence of equality, which allows us to pass
from a family to the corresponding predicate, their algebraic structure is the
same as that of predicates.  However, if we abstain from use of equality, the
situation is as follows.  The construction of set indexed suprema can be
carried through
\be
  \bigcup_{i\in I} \FAM{ f_i(t) }{ t \in T_i } &==& \FAM{ f_i(t) }{ <i,t> \in \SI{i\in I} T_i } \ \hbox{,}
\ee
which gives a sup-lattice.  Additionally for any $s \in S$ we can form the
singleton family $\FAM{ s }{ i \in I }$ taking for $I$ any non-empty set.

We cannot \emph{say} that an element of $S$ belongs to a family $\FAM{ f(i) }{
i \in I }$.  Still less can we say that one family includes another, or
overlaps with it (as this requires an equation).  What we \emph{can} state
however is that a family is included in a predicate, or that it overlaps with
it:
\be
  \FAM{ f(i) }{ i \in I } \sub U &==& (\forall i \in I)\,U\big(f(i)\big) \ \hbox{;}\\
  \FAM{ f(i) }{ i \in I } \olp U &==& (\exists i \in I)\,U\big(f(i)\big) \ \hbox{.}
\ee


\smallbreak
To summarize, predicates have a rich algebraic structure.  In contrast, the
structure of families is impoverished, supporting only suprema operations of
various kinds.  To compensate, we have a concrete, computational form of the
notion of subset.

\subsection{The general notion of binary relation}

A binary relation between two sets $S_1$ and $S_2$ is a subset of the
cartesian product $S_1×S_2$, or to put it another way, a function from $S_1$
to subsets of $S_2$:
\be
 \Pow(S_1×S_2) & =      & (S_1×S_2) -> \Set    \\
               & \simeq & S_1 -> (S_2 -> \Set) \\
               & =      & S_1 -> \Pow(S_2)     \ \hbox{.}
\ee
We will leave implicit the isomorphism (``currying'') between the two
versions.  There are thus two ways to write ``$s_1$ and $s_2$ are related
through $R\sub S_1×S_2$'': either ``$(s_1,s_2) |= R$'' or ``$s_2 |= R(s_1)$''.

\smallbreak
Because relations are subset valued functions, they inherit all the algebraic
structure of predicates pointwise.  Additionally, we can define the following
operations.

\begin{description}

\item[Converse:]
\RULE{R \sub  S_1 × S_2}%
     {\CONV{R} \sub S_2 × S_1}
with $(s_2,s_1)|=\CONV{R} == (s_1,s_2)|=R$~.

\smallbreak
\item[Equality:] $\eq \sub S×S$ with $\eq(s) = \SING{s}$~.  \COMMENT{(This
  requires equality!)}

\smallbreak
\item[Composition:]
\RULE{Q \sub S_1×S_2  &  R\sub S_2×S_3}%
     { Q \SEQ R \sub S_1×S_3}\\
with
$ (s_1,s_3) |= (Q\SEQ R) == (\exists s_2\in S_2)\ (s_1,s_2)|= Q \,\hbox{and}\,(s_2,s_3)|= R$~.

\smallbreak
\item[Reflexive and transitive closure:]
\RULE{R \sub S×S}%
     {\RTC{R} \sub S×S}\\
with $\RTC{R} == \eq \cup R \cup (R\SEQ R) \cup R^3 \cup \ldots$
\kern1cm\COMMENT{(inductive definition)}

\noindent
Note that the ``reflexive'' part requires equality to be definable.

\smallbreak
\item[Post and pre-division:]\leavevmode
  \begin{itemize}
    \item
      \RULE{ Q \sub S_1×S_3 &   R \sub S_2×S_3}%
           {(Q \CHOP R) \sub S_1×S_2}\\
      with $(s_1,s_2) |= (Q \CHOP R) == R(s_2) \sub Q(s_1)$~;

    \item
      \RULE{ Q \sub S_1×S_3  &  R \sub S_1×S_2}
           {(R \CHIP Q) \sub S_2×S_3}
      with $(R \CHIP Q) == \CONV{(\CONV{Q} / \CONV{R})}$~.
\end{itemize}

\end{description}
These operators satisfy a wealth of familiar algebraic laws, from which we
want to recall only the following.

\begin{itemize}

\item Composition and equality are the operators of a monoid. Composition is
  monotone in both arguments, and in fact commutes with arbitrary unions on
  both sides.

\item Post-composition $(\BLANK\SEQ R)$ is left-adjoint to post-division
  $(\BLANK\CHOP R)$; similarly, pre-composition $(R \SEQ\BLANK)$ is
  left-adjoint to pre-division $(R \CHIP\BLANK)$.

\item Converse is involutive and reverses composition: $\CONV{(Q \SEQ R)} =
  \CONV{R} \SEQ \CONV{Q}$.

\item For each function $f \in S_1 -> S_2$, its graph relation $\gr f \sub S_1
  × S_2$ satisfies both $\eq_{S_1} \sub (\gr f) \SEQ \CONV{(\gr f)}$
  (totality), and $\CONV{(\gr f)} \SEQ \gr f \sub \eq_{S_2}$ (determinacy).

\end{itemize}

\subsection{Transition structures}

What happens to the notion of a binary relation if we replace the
contravariant functor $\Pow(\BLANK)$ with the co-variant functor
$\Fam(\BLANK)$?  This gives two candidates for a computational representation
of relations:
\be
  \Fam(S_1 \times S_2)   \qquad  \hbox{and}\qquad
  S_1 -> \Fam(S_2)\ \hbox{.}
\ee

\begin{itemize}

\item In more detail, an object of the first type consists of a set $I$,
  together with a pair of functions with $I$ as their common domain: $f \in I
  -> S_1$ and $g \in I -> S_2$.  Such a pair is commonly known as a
  \newterm{span}.

\item On the other hand, an object $T$ of the second type consists of a
  function~$F$ which assigns to each $s \in S_1$ a family of $S_2$'s, that we
  may write
  \be
    F(s) = \FAM{ n(s,t) }{ t \in A(s) } \ \hbox{.}
  \ee
  Where $A : S_1 -> Set$ and $n \in \PI{s_1\in S_1} A(s_1) -> S_2$.  We call
  this a \newterm{transition structure}. When no confusion arises, we write
  ``$s[a]$'' instead of ``$n(s,a)$''.

\end{itemize}

In contrast with the situation with relations, any isomorphism that can be
defined between spans and transition structures seems to require use of an
equality relation.  Transition structures are inherently asymmetric.  There is
a genuine bifurcation between spans and transition structures.  In this paper
we shall be concerned only with transition structures.  To some extent, the
relationship between spans and transition structures remains to be explored.

\medbreak
Transition structures sometimes provide a more appropriate model than
relations for ``asymmetric'' situations in which one of the terms of the
relation has priority or precedence in some sense.
\begin{itemize}

\item The notion of an occurrence of a subexpression of a first-order
  expression can be represented by a transition structure on expressions, in
  which the set $A(s)$ represents the set of positions within $s$, and $s[a]$
  represents the subexpression of $s$ that occurs at position $a$.

\item In general rewriting systems, an expression is rewritten according to a
  given set of rewriting rules.
  In state $s$, each rule can be represented by an $a\in A(s)$, where $s[a]$
  is the result of the rewriting of $s$ by the rule $a$.

\item A deterministic automaton that reads a stream of characters, changing
  state in response to successive characters can be represented by a
  transition structure.  In such a case, one usually writes $s \transition{a}
  s'$ for $s[a]=s'$.

\end{itemize}

In comparison with relations, transition structures have weaker algebraic
properties.  There are transition structure representations for equality
relations and more generally the graphs of functions, and for indexed unions,
composition, and closure operations such as reflexive and transitive closure:
transition structures form a Kleene algebra.
\begin{description}
\item[Composition:]
  \RULE{T_1 : S_1 -> \Fam(S_2) & T_2 : S_2 -> \Fam(S_3)}
       {(T_1 \SEQ T_2) : S_1 -> \Fam(S_3)}

  where the components $\TSI{(T_1\SEQ T_2)}$ and $\TSn{(T_1\SEQ T_2)}$ of
  $T_1\SEQ T_2$ are defined as:
   \be
     \TSI{(T_1\SEQ T_2)}(s_1) &==& \SI{t_1 : \TSI{T_1}(s_1)} \TSI{T_2}(s_1[t_1])\\
     \TSn{(T_1\SEQ T_2)}<t_1,t_2>         &==& (s_1[t_1])[t_2] \ \hbox{.}
   \ee

\item[Identity:]
  $\eq : S -> \Fam(S)$ with $T(\BLANK) == \One$ and $s[\BLANK] == s$.
  Note that the equality relation is not necessary to define this interaction
  structure.

\end{description}
The definitions are straightforward, and the reader is encouraged to try the
case of reflexive and transitive closure for themselves.

\smallbreak
On the other hand, transition structures are not closed under intersection,
converse, or division.  They can however be used as pre-components to
relations, and as post-divisors of relations.  The definitions, which make no
use of equality, are as follows.
\be
  (s_1,s_3) |= (T \SEQ R)    &==&    T(s_1) \olp \CONV{R}(s_3)\ \hbox{;}  \\
  (s_1,s_2) |= (R \CHOP T)   &==&    T(s_2) \sub R(s_1) \ \hbox{.}
\ee
(In the first equation, $T : S_1 -> \Fam(S_2)$ and $R \sub S_2×S_3$, while in
the second, $R \sub S_1 × S_3$ and $T : S_2 -> \Fam(S_3)$.)

\smallbreak
Note that we can define the relation corresponding to a transition structure
by precomposing the transition structure to equality: if $T:S_1 -> \Fam(S_2)$,
define $T^\circ : S_1 -> \Pow(S_2)$ as $T\SEQ\eq_{S_2}$.


\section{Predicate transformers} \label{sec:PT}

\subsection{Motivations and basic definitions}

A predicate transformer is a function from subsets of one set to subsets of
another:
\be
  \Pow(S_2) ->\Pow(S_1) &=&      \Pow(S_2) -> S_1 -> \Set           \\
                        &\simeq& \big(\Pow(S_2) × S_1\big) -> \Set \\
                        &\simeq& \big(S_1×\Pow(S_2)\big) -> \Set   \\
                        &\simeq& S_1 -> \big(\Pow(S_2)->\Set\big)  \\
                        &=& S_1 ->\Pow\big(\Pow(S_2)\big)     \ \hbox{.}
\ee
As these isomorphisms show, from another point of view, a predicate
transformer is nothing but a higher-order relation (between elements of one
set and subsets of another).

\medbreak
Since the mid-70's, predicate transformers have been used as denotations for
commands such as assignment statements in imperative programming languages.
Some predicate transformers commonly considered in computer science are the
weakest precondition operator, the weakest liberal precondition, the strongest
postcondition (all introduced by Dijkstra), and the weakest and strongest
invariant of a concurrent program (introduced by Lamport).  Perhaps the most
fundamental of these is the weakest precondition.  In weakest precondition
semantics, one associates to a program statement~$P$ a predicate
transformer~$|P|$ mapping a \emph{goal} predicate (which one would like to
bring about) to an \emph{initial} predicate (which ensures that execution of
$P$ terminates in a state satisfying the goal predicate).  On the other hand,
the weakest liberal precondition is more relevant in connection with
predicates which one would like to avoid or maintain.

\smallbreak
In an effort to cut down the semantic domain of predicate transformers to
those that are in some sense executable, various ``healthiness''
properties\footnote{like strictness, distribution over intersections,
distribution over directed unions} have been required of predicate
transformers. In the 80's and 90's reasons emerged for relaxing most
such restrictions, except for the most basic, monotonicity.  In explanation of
monotonicity, if a goal predicate is weakened (made easier to achieve), the
corresponding initial predicate should be weakened.  More technically, the
Knaster-Tarski theorem is heavily exploited in developing the semantics of
recursion and iteration. In the following, the qualification ``monotone'' will
be implicit: all predicate transformers will be monotone, except where
explicitly indicated.

An active field of computer science instigated by Morgan, Morris, Back, and
von Wright is now founded on the use of monotone predicate transformers not
just as a semantic domain for commands, but as a framework for developing
imperative programs from specifications.  This field is called ``refinement
calculus''; the canonical reference for the refinement calculus is Back and
von Wright's textbook \cite{BvWRC}.

The refinement calculus is a ``wide spectrum'' language in the sense that both
programs and specifications are represented by monotone predicate
transformers. (In contrast, in type theory programs and specifications lie,
roughly speaking, on opposite sides of the ``$\in$'' symbol.) Specifications
are manipulated into an executable form (acquiring various healthiness
conditions), until they can be coded in a real programming notation.


\subsection{Algebraic structure}

The lattice structure of predicates lifts pointwise to the level of relations.
Analogously, the lattice structure lifts to the level of predicate
transformers:

\begin{itemize}
  \item predicate transformers are ordered by pointwise inclusion:
    \be
      F \sub G   &==& \hbox{\textbf{``}}(\forall U \sub S)\ F(U) \sub G(U)\hbox{\textbf{''}} \ \hbox{;}
    \ee
    \ie~``$F \sub G$'' is a shorthand for the \emph{judgment} ``$U\sub S |-
    F(U) \sub G(U)$'' and is not an actual proposition or set.
  \item they are closed under intersection and union:
    \be
      \big(\bigcup_i F_i\big)(U)  &==&  \bigcup_i \big(F_i(U)\big) \ \hbox{;}\\
      \big(\bigcap_i F_i\big)(U)  &==&  \bigcap_i \big(F_i(U)\big) \ \hbox{.}
    \ee
\end{itemize}

The bottom and top of the lattice are conventionally called $\ABORT$ and
$\MAGIC$ respectively.  The predicate transformer~$\ABORT$ transforms all
predicates to the empty predicate: it is impossible to achieve anything by use
of a resource satisfying $\ABORT$. On the other hand, $\MAGIC$ transforms all
predicates to the trivial predicate, which always holds.  A resource
fulfilling the $\MAGIC$ specification could be used to accomplish anything,
even the impossible.

\medbreak
Just as relations support not only a lattice structure, but also a monoidal
structure of composition, so it is with predicate transformers.  Predicate
transformers are of course closed under composition:
\[
  F \SEQ G  ==  F \cdot G \ \hbox{;}
\]
and the unit of composition is conventionally called \SKIP:
\[
  \SKIP(U) == U \  \hbox{.}
\]
Both relational and predicate transformer compositions are monotone.  The
distributivity laws satisfied by ``$\SEQ$'' are however quite different from
the case of relations.  With relations, composition distributes over unions on
both sides, though not (in general) over intersections.  With predicate
transformers, composition distributes over both intersections and unions on
the left, though not in general over either intersection or union on the
right.

\subsection{Angelic and demonic update}

Somewhat as a function $f \in S_1 -> S_2$ lifts to a relation $(\gr f) : S_1
->\Pow(S_2)$, so a relation $R : S_1 ->\Pow(S_2)$ lifts to a predicate
transformer.  However in this case there are two lift operations.  These are
conventionally called angelic and demonic update.
\begin{center}
  \RULE{R : S_1 ->\Pow(S_2)}%
       {\angelU{R},\demonU{R} : \Pow(S_2) -> \Pow(S_1)}
\end{center}
with:\footnote{Note that we have diverged slightly from the notation of Back
and von Wright.  In their notation, the angelic update $\angelU{R}$ is written
$\{R\}$.}
\be
  \angelU{R}(U) &==& \SET{ s_1 \in S_1 }{ R(s_1) \olp U } \ \hbox{;}& \COMMENT{(angelic update)} \\
  \demonU{R}(U) &==& \SET{ s_1 \in S_1 }{ R(s_1) \sub U } \ \hbox{.}& \COMMENT{(demonic update)}
\ee

Note also that $\angelU{\CONV{R}}(U)$ is nothing but the set of states related
by $\CONV{R}$ to states that satisfy $U$ or, in other words, the direct
relational image of $U$ under $R$.  When there is no danger of confusion, we
shall in the following write $R(U)$ for $\angelU{\CONV{R}}(U)$ and $R$ for
$\angelU{\CONV{R}}$.

\medbreak

At first sight, the angelic and demonic updates may look a little strange.
What do they have to do with programming?  In two particular cases though,
they are immediately recognizable, namely when firstly, the relation is
included in the equality relation on a state-space; and secondly when the
relation is the graph of a function.

\begin{description}

\item[Assertions and assumptions:]
  when the relation $R$ is a subset of the identity relation (which can be
  identified with a predicate $U$), the angelic update $\angelU{U}$ is known
  as an \newterm{assertion} (that the Angel is obliged to prove), whereas the
  demonic update $\demonU{U}$ is known as an \newterm{assumption} (that the
  Demon is obliged to prove).  Assertion and assumption satisfy the
  equivalences:
  \be
    \assert{U}(V) &=& U \cap V \qquad \hbox{and} \qquad
    \assume{U}(V) &=&  \SET{s \in S}{U(s) \rightarrow V(s)}  \ \hbox{.}
  \ee

\item[Assignments:]
  because singleton predicates~$\SING{s}$ satisfy the equivalences
  \be
    \SING{s} \olp U &<=>& s |= U &<=>& \SING{s} \sub U \ \hbox{,}
  \ee

  it follows that if $f \in S_1 -> S_2$, we have $\angelU{\gr f}(U) \simeq U·f
  \simeq \demonU{\gr f}(U)$.  In this case the predicate transformer commutes
  with arbitrary intersections and unions.  The canonical example of such an
  update is the assignment statement $x := e$ where $x$ is a state variable,
  and $e$ is a ``side-effect free'' mathematical expression that may refer to
  the values of other state variables.  This is interpreted as the
  ``substitution'' predicate transformer $U |-> \SET{s \in S}{f(s) |= U }$,
  where $f \in S -> S$ is the function that maps a state $s$ to the state $s'$
  in which all variables except $x$ have the same value as in $s$, and the
  value of $x$ in $s'$ is the denotation of the expression $e$ in state
  $s$.\footnote{It would take us too far afield to fully explain the syntax
  and semantics of state variables and assignment statements.}

\end{description}

\subsection{Fundamental adjunction}

Perhaps the most fundamental law in the refinement calculus, with the same
pivotal r\^ole as Sambin's ``fundamental adjunction'' in his development of
basic topology through basic pairs (\cite{BPI}) is the following Galois
connection between angelic and demonic updates.

\begin{prop}\label{prop:fundamental-adjunction}
Suppose $R \sub S_1×S_2$;  we have, for all $U \sub S_1$,
$V\sub S_2$
\[
  \angelU{\CONV{R}}(U) \sub V \quad <=>\quad  U \sub \demonU{R}(V)\ \hbox{,}
\]
which is commonly written $\angelU{\CONV{R}} \dashv \demonU{R}$.
\end{prop}
\begin{proof}
  Straightforward.
\end{proof}
Points \textit{1} and \textit{2} of the following corollary are the ground for
all the development of basic topology from ``basic pairs'' (\cite{BPI}).
Recall that an interior [closure] operator is a predicate
  transformer $P$ satisfying:
  \be
    \hbox{\textit{closure}}       &       & \hbox{\textit{interior}}\\
    U \sub P(U)                   & \quad & P(U) \sub U \\
    U \sub P(V) => P(U) \sub P(V) &       & P(U) \sub V => P(U)\sub P(V)
  \ee
\begin{cor} \label{cor:gc-consequences}
We have:
\begin{enumerate}
\item $  \angelU{\CONV{R}}\SEQ\demonU{R}$ is an interior operator, in
  particular:  $  \angelU{\CONV{R}} \SEQ \demonU{R} \sub \SKIP  $;
\item $  \demonU{R}\SEQ\angelU{\CONV{R}}$ is a closure operator, in
  particular: $  \SKIP  \sub  \demonU{R} \SEQ \angelU{\CONV{R}}  $;
\item $\demonU{R}  = \demonU{R} \SEQ \angelU{\CONV{R}} \SEQ \demonU{R}$ and
  $\angelU{\CONV{R}} = \angelU{\CONV{R}} \SEQ \demonU{R} \SEQ \angelU{\CONV{R}}$;
\item $\angelU{\CONV{R}}$ commutes with all unions and $\demonU{R}$ commutes
  with all intersections.
\end{enumerate}
\end{cor}
\begin{proof}
  Straightforward. 
\end{proof}

\medbreak
Back and von Wright's textbook on the refinement calculus contains many normal
form theorems that relate the properties of a predicate transformer to its
expression in the refinement calculus.  Among these, the most general is the
following.  It provides one motivation for the analysis of predicate
transformers given in section~\ref{sec:IS} below.
\begin{quotation} \label{normalFormTh}
  \noindent
  \textbf{Theorem 13.10.}  Let $S$ be an arbitrary monotonic predicate
  transformer term.  Then there exist state relation terms $P$ and $Q$ such
  that $S = \angelU{P} \SEQ \demonU{Q}$. \kern5mm (\cite{BvWRC}, p. 220, with
  $\{Q\}$ changed to $\angelU{Q}$)\footnote{The proof given is an manipulation
  in higher-order logic, in which the relation $Q$ is taken to be the
  membership relation.}
\end{quotation}
In other words, so far as monotone predicate transformers are concerned, it
suffices to consider those in which an angelic update is followed by a demonic
update.  In section~\ref{sec:IS}, we will represent predicate transformers by
such a composition, where the update relations are each given by transition
structures.

\subsection{Iterative constructions}

The most interesting construct are connected with iteration.
(One of the main applications of our category will be to model iterative
client-server interaction, in section~\ref{sec:clientServer}.)

In the case of relations and transition structures, 
there is a single notion of iteration, namely the reflexive and transitive closure. 
However in the case of predicate transformers, there are two different iteration
operators: one orientated toward the Angel, and the other toward the Demon.

According to the Knaster-Tarski theorem, each monotone predicate transformer
$F :\Pow(S) ->\Pow(S)$ possesses both a least fixpoint $\mu F$ and a
greatest fixpoint $\nu F$. They can be defined as:
\be
  \MU{X} F(X) &==& \bigcap \SET{U \sub S}{ F(U) \sub U } :\Pow(S)\ \hbox{;} \\
  \NU{X} F(X) &==& \bigcup \SET{U \sub S}{ U \sub F(U) } :\Pow(S)\ \hbox{.}
\ee
Note that the intersection and union operators are applied to a higher order
predicate (a predicate of predicates, rather than a family of predicates).  In
a predicative framework we therefore run into difficult questions about the
justification of those very general forms of induction and coinduction.  In
this paper we attempt no answer to these foundational questions: we need to
consider only certain forms of ``tail'' recursion, in which the $\mu$- or
$\nu$-bound variable occurs only as the right-hand operand of $\SEQ$.

The two operations we need are written $\RTC{\BLANK}$ and $\FISH{\BLANK}$, and
are characterized by the laws:\footnote{Yet again we diverge from (and indeed
clash with) the usual notation of Back and von Wright's refinement calculus.
What we call angelic iteration, and write \RTC{F} is written there
$F^\emptyset$ (and also called angelic iteration).  What we call demonic
iteration and write \FISH{F} is written there $F^\ast$, and called weak
iteration.}

\begin{center}
\begin{tabular}{ll}
  \RULE{ F : \Pow(S) -> \Pow(S)}%
       {\RTC{F},\FISH{F} : \Pow(S) -> \Pow(S)}
\end{tabular}
\end{center}
with the rules:
\be
  \SKIP \cup (F \SEQ \RTC{F})  \sub  \RTC{F}
  &\hbox{,}&
  \hbox{\RULE{\SKIP \cup (F \SEQ G) \sub G}{\RTC{F} \sub G}} \ \hbox{;}\\
  \FISH{F}  \sub  \SKIP \cap (F \SEQ \FISH{F})
  &\hbox{,}&
  \hbox{\RULE{G \sub \SKIP \cap (F \SEQ G)}{G \sub \FISH{F}}} \ \hbox{.}
\ee

\smallbreak
\noindent
We may define these operations using $\mu$ and $\nu$ as:
\be
  \RTC{F}(U)      &==& \MU{X} U \cup (F \SEQ X) \qquad \hbox{and}\qquad
  \FISH{F}(V)     &==& \NU{X} V \cap (F \SEQ X) \ \hbox{.}
\ee

Both are iterative constructions.  In the case of $\RTC{F}$ the iteration must
be finite and the Angel chooses when to exit.  In the case of $\FISH{F}$, the
iteration may be infinite, and the Demon chooses when (if ever) to exit.

\begin{prop} \label{prop:FISH_RTC_int_cl}
  If $F$ is a predicate transformer, then $\RTC{F}$ is a closure operator and
  $\FISH{F}$ is an interior operator.
\end{prop}
\begin{proof}
  We give only the proof that $\RTC{F}$ is a closure operator. The
  proof that $\FISH{F}$ is an interior operator is completely dual.
  \begin{itemize}

    \item $U \sub \RTC{F}(U)$: we know that $\RTC{F}(U)$ is a pre-fixpoint
      of $X |-> U \cup F(X)$, which means that $U\cup F\big(\RTC{F}(U)\big)
      \sub \RTC{F}(U)$, and so $U\sub\RTC{F}(U)$.

    \item $U \sub \RTC{F}(V) => \RTC{F}(U) \sub \RTC{F}(V)$.  Suppose that $U
      \sub \RTC{F}(V)$.  Since $\RTC{F}(U)$ is the least pre-fixpoint of $X
      |-> U \cup F(X)$, it suffices to show that $\RTC{F}(V)$ is also a
      pre-fixpoint of this operator, \ie~that $U\cup F\big(\RTC{F}(V)\big)
      \sub \RTC{F}(V)$. Since $\RTC{F}(V)$ is a pre-fixpoint for $X|->V\cup
      F(X)$, we have $F\big(\RTC{F}(V)\big) \sub \RTC{F}(V)$, and by
      hypothesis, we have $U\sub\RTC{F}(V)$.  We can conclude.
  \end{itemize}
It is worth noting that the operation $\RTC{}$ itself is a closure operation
on the lattice of predicate transformers, but that $\FISH{}$ is \emph{not} an
interior operator.

\end{proof}
Some other properties of those operations are given by the following lemma.
First, a definition:
\begin{defn}\label{defn:invariant-saturated}
  Suppose $F$ is a predicate transformer.
  \begin{enumerate}

    \item an \newterm{$F$-invariant}, (or simply an \newterm{invariant} when
      $F$ is clear) is a post-fixpoint of $F$, \ie~a predicate $U$
      satisfying $U \sub F(U)$;

    \item an \newterm{$F$-saturated} predicate, (or simply a
      \newterm{saturated} predicate when $F$ is clear) is a pre-fixpoint of
      $F$, \ie~a predicate $U$ satisfying $F(U) \sub U$.

  \end{enumerate}
\end{defn}

\noindent
We have:
\begin{lem} \label{lem:invFISH_satRTC}
  If $F$ is a predicate transformer on $S$ and $U\sub S$, we have:
  \begin{itemize}
    \item $\RTC{F}(U)$ is the strongest (\ie~least) $F$-saturated predicate including $U$;
    \item $\FISH{F}(U)$ is the weakest (\ie~greatest) $F$-invariant contained in $U$.
  \end{itemize}
\end{lem}
\begin{proof}
  We will prove only the second point, as the first one is completely dual.
  \begin{itemize}
    \item $\FISH{F}(U)$ is contained in $U$: this is a consequence of
      $\FISH{F}$ being an interior operator.
      (Proposition~\ref{prop:FISH_RTC_int_cl}.)

    \item $\FISH{F}(U)$ is $F$-invariant: $\FISH{F}(U)$ is the greatest
      post-fixpoint of the operator $X |-> U \cap F(X)$; in particular,
      $\FISH{F}(U) \sub U \cap F\big(\FISH{F}(U)\big)$, which implies that
      $\FISH{F}(U)\sub F\big(\FISH{F}(U)\big)$.

    \item $\FISH{F}(U)$ is the greatest such invariant: suppose that $V$ is
      another invariant contained in $U$, \ie~we have $V \sub F(V)$ and $V\sub
      U$. This implies that $V$ is a post-fixpoint of the above operator.
      Since $\FISH{F}(U)$ is the greatest post-fixpoint, we conclude directly
      that $V\sub\FISH{F}(U)$.
  \end{itemize}
\end{proof}


\section{Interaction structures} \label{sec:IS}

\subsection{Motivations}

As in the case of relations, we obtain another more computationally oriented
notion of predicate transformer by replacing the $\Pow(\BLANK)$ functor with
the $\Fam(\BLANK)$ functor.  There is again more than one way to do this.  We
will focus on the structure arising from the representation of
predicate transformers as $S -> \Pow^2(S')$:
\[
  w : S -> \Fam^2(S') \ \hbox{.}
\]
Expanding the definition of $\Fam(\BLANK)$, we see that the declaration of $w$
consists of the following data:
\begin{enumerate}
  \item a function  $A : S -> \Set$;                                
  \item a function  $D : \PI{ s \in S }\ A(s) -> \Set$;             
  \item a function  $n : \PI{ s \in S, a \in A(s)}\ D(s,a) -> S'$.  
\end{enumerate}

In essentials, the invention of this structure should be attributed to
Petersson and Synek (though similar constructions were implicitly present in
earlier works:~\cite{tarski,grzeg,dragalin}). In \cite{PS89}, they introduced
a set-constructor for a certain inductively defined family of trees, relative
to the signature
\be
       A : \Set                                                   \\
       B(x) : \Set     &  \WHERE{ x \in A }                       \\
       C(x,y) : \Set   &  \WHERE{ x \in A, y \in B(x)}            \\
       d(x,y,z) \in A  &  \WHERE{ x \in A, y \in B(x), z \in C(x,y) }
\ee
which is nothing more than a pair of a set $A$ and an element of $A ->
\Fam^2(A)$.
We will make use of (a slight variant of) their definitional schema in
defining one of our iteration operators below,
namely ``angelic iteration''.

\subsection{Applications of interaction structures}

This type is rich in applications.  Broadly speaking these applications fall
under two headings: interaction and inference.

\paragraph{Interaction.}

This is our main application.
\begin{itemize}
  \item[$S$:] we take $S$ to model the state space of a device.
    We prefer to call this the state of the \emph{interface} as the device
    itself may have a complicated internal state which we need not
    understand to make use of the device.  For example, think of $s \in S$
    as the state of one's bank balance, as it is observed by someone using
    an ATM.\footnote{Automatic Teller Machine ---a cash machine.}

  \item[$A$:] for each state $s \in S$, we take the set $A(s)$ to model
    the set of commands that the user may issue to the device.
    For example, think of $a \in A(s)$ as a request to withdraw cash from an
    ATM.

  \item[$D$:] For each $s \in S$ and $a \in A(s)$,
    we take $D(s,a)$ to model the set of responses that the device
    may return to the command~$a$.
    It is possible that there is more than one response that the device may legitimately return
    to a given command.
    For example, think of the response \texttt{Service Unavailable} to a
    withdrawal request.

  \item[$n$:] For each $s \in S$, command $a \in A(s)$ and response $d
    \in D(s,a)$, we take $n(s,a,d)$ to model the next state of the
    interface.  Note that the next state is \emph{determined} by the response.
    This means that the current state of the system can always be computed from 
    its initial state, together with a complete record of commands and
    responses exchanged \textit{ab initio}.
\end{itemize}
The two agents interacting across such a command-response interface are
conventionally called the Angel (for a pronoun we use ``she''), and the Demon
(``he'').  The Angel issues commands and receives responses.  She is active,
in that she has the initiative in any interaction.  The Demon is passive, and
merely obeys instructions, to each of which he returns a response.  The
terminology of Angels and Demons is rife in the refinement calculus
literature, in which an interface is thought of as a contract regulating the
behavior of two parties, the Angel and Demon. 
We have named the two components of an interaction structure $A$ and $D$ after
them. 
(Alternative \textit{dramatis personae} might be $\exists$loise and $\forall$belard, Opponent and
Defendant, Master and Slave, Client and Server.)


Other applications that have broadly the same interactive character are
indicated in the following table.
\begin{center}
  \footnotesize
  \begin{tabular}[t]{|l|l|l|l|l|}
    \hline
    \textit{idiom}& $S$            & $A$             & $D$             &   $n$\\
    \hline
    game          & state          & moves           & counter-moves   &   next state \\
    system        & state          & system call     & return          &   next state \\
    experiment    & knowledge      & stimulus        & response        & \\
    examination   & knowledge      &  question       & answer          & \\
    \hline
  \end{tabular}
\end{center}

\paragraph{Inference.}

A second style of application of the structure (which plays no explicit
r\^ole in this paper) is to model an inference system, or (to use Aczel's
term) a rule-set.  One does not attempt here to capture the idea of a
schematic rule, but rather the inference steps that are instances of such
rules.
\begin{itemize}
\item[$S$:] we may take the elements of the set $S$ to model judgments that
  can stand `positively' at the conclusion or occur `negatively' as some premise of an inference step.

\item[$A$:] for each judgment $s \in S$, we may take the elements of the set
  $A(s)$ to model inference steps with conclusion $s$.
  
\item[$D$:] for each judgment $s \in S$ and inference step $a \in
  A(s)$ by which $s$ can be inferred, we may take the elements of the
  set $D(s,a)$ to index, locate, or identify one of the family of premises
  required to infer $s$ by inference step $a$.

 \item[$n$:] for each judgment $s \in A$, inference step $a \in A(s)$,
   and index $d \in D(s,a)$ for a premise of that inference step, we
   may take $n(s,a,d)$ to model the judgment to be proved at premise $d$ in
   inference step $a$.
\end{itemize}
Instead of judgements and inference steps, we may consider grammatical categories and 
productions as in Petersson and Synek's original application (\cite{PS89}), or
sorts and (multi-sorted) signatures.


\subsection{Definition and basic properties}

\begin{defn}
  If $S$ and $S'$ are sets, an object $w$ of type
       $S -> \Fam^2(S')$
  is called an \newterm{interaction structure} (from $S$ to $S'$).  We refer to the
  components of $w$ as follows:
  \be
    \ISA{w} &:& S -> \Set                                               \\
    \ISD{w} &:& \PI{s \in S}\ \ISA{w}(s) -> \Set                        \\
    \ISn{w} &\in& \PI{s \in S,\, a \in \ISA{w}(s)}\  \ISD{w}(s,a) -> S'
  \ee
  When no confusion is possible, we prefer to leave the ``$w.$'' implicit, and
  simply write $A$, $D$ and $n$, possibly with decorations. We also use the
  notation $s[a/d]$ as a synonym for $\ISn{w}(a,s,d)$ when $w$ is clear from
  the context.
\end{defn}

Before examining the objects of this type in more detail, we mention some
other representations of predicate transformers:
\begin{itemize} \label{discussionAlternativeIS}

  \item $S_1 -> \Pow^2(S_2) \simeq \Pow(S_2)->\Pow(S_1)$:  this is the notion
    studied in section~\ref{sec:PT}, or in Back and von Wright's book
    (\cite[sec. 5.1, p. 251]{BvWRC}) under the name ``choice semantics'';

  \item $S_1 -> \Pow\big(\Fam(S_2)\big) \simeq \Fam(S_2) -> \Pow(S_1)$: this notion is
    very similar to the previous one (they are equivalent in the presence of
    equality). To our knowledge, this type has never been considered as a
    viable notion;

  \item $S_1 -> \Fam\big(\Pow(S_2)\big)$: because a subset on a proper type need not
    be equivalent to a set indexed family on the same type, this notion is
    intrinsically different from the previous two. This is the notion used by
    Aczel to model generalized inductive definitions in~\cite{aczel}.
    This is also the structure used in~\cite{induc_top} under the name
    \newterm{axiom set}.

    From our perspective, this notion seems to abstract away the action of the
    Demon: the Angel doesn't see the Demon's reaction, but only a property of
    the state it produces.  The Demon's reaction is in some sense ``hidden''.
\end{itemize}
There are other variants based on types isomorphic to $\Pow(S_2) -> \Pow(S_1)$
such as $\Fam(S_2) -> \Fam(S_1)$, $\Pow(S_2 × \Fam(S_1))$ and so on.  We have
not investigated all the possibilities systematically, but none of them seems
to fit our purpose.

\bigbreak
Associated with an interaction structure $w$ from $S$ to $S'$ are two monotone
predicate transformers  $w^\circ$ and $w^\bullet :\Pow(S') ->\Pow(S) $.  Both
are concerned with the notion of ``reachability'' of a predicate (on $S'$)
from a state (in $S$). The difference is which agent tries to bring about the
predicate: either the Angel (in the case of $w^\circ$) or the Demon (in the
case of $w^\bullet$).
\begin{defn}
  If $w=<A,D,n>$ is an interaction structure on $S$, define:\\
  {\scriptsize(recall that ``\/$\exists$'' and ``\/$\forall$'' are synonyms for
  ``\/$\Sigma$'' and ``\/$\Pi$'')}
  \be
    s|=w^\circ(U)   &<=>& \big(\exists a\in A(s)\big)\,\big(\forall d\in D(s,a)\big)\ s[a/d]|=U \ \hbox{;}\\
    s|=w^\bullet(U) &<=>& \big(\forall a\in A(s)\big)\,\big(\exists d\in D(s,a)\big)\ s[a/d]|=U \ \hbox{.}\\
  \ee
\end{defn}

\smallbreak
Of these, lemma~\ref{lem:ISDual} below shows that $\BLANK^\circ$ is more
fundamental.

\begin{defn}  \label{def:ISDual}
  If $S$ and $S'$ are sets, and $w$ is an interaction structure from $S$ to $S'$,
  define $w^\bot : S -> \Fam^2(S')$ as follows.
  \be
    \ISA{w^\bot}(s)          &==& \PI{ a \in \ISA{w}(s) }\ \ISD{w}(s,a) \\
    \ISD{w^\bot}(s,\BLANK)   &==& \ISA{w}(s) \\
    \ISn{w^\bot}(s,f,a)      &==& s[a/f(a)]
  \ee
\end{defn}

\noindent
As we'll see in proposition~\ref{prop:ISPT}, this is a constructive version of
the dual operator on predicate transformers.  Although this operation doesn't
enjoy all the duality properties of its classical version (in particular, it
is not provably involutive), we still have the following:
\begin{lem} \label{lem:ISDual}
  For any interaction structure $w$, we have: $w^\bullet  =  (w^\bot)^\circ$.
\end{lem}
\begin{proof}
 Axiom of choice.
\end{proof}
\noindent
The converse $w^\circ=(w^\bot)^\bullet$ holds classically but not constructively.

\subsubsection{Lattice structure, monoidal operations}

We define inclusion between interaction structures by interpreting them as
predicate transformers via the ${}^\circ$ operator:
\begin{defn}
  Define $w_1 \sub w_2   ==   {w_1}^\circ \sub {w_2}^\circ$.
\end{defn}
\noindent
Once again, this is not a proposition, but only a judgment.

In contrast with the impoverished structure of transition structures relative
to relations, interaction structures support
the full algebraic structure of monotone predicate transformers, as we now
show.

\begin{defn}
  Define the following operations on interaction structures:

  \begin{description}

  \item[Updates]
      If $T=<A,n> : S -> \Fam(S')$ is a transition structure, then
        \be
          \ISA{\angelU{T}}(s)          &==& A(s)        & \qquad & \ISA{\demonU{T}}(s)          &==& \One \\
          \ISD{\angelU{T}}(s,a)        &==& \One        & \qquad & \ISD{\demonU{T}}(s,\BLANK)   &==& A(s)\\
          \ISn{\angelU{T}}(s,a,\BLANK) &==& s[a]\quad   & \qquad & \ISn{\demonU{T}}(s,\BLANK,a) &==& s[a] \ \hbox{.}
        \ee

  \item[Extrema]
      If $(w_i)_{i \in I}$ is an indexed family of interaction
      structures, then
      \be
      \ISA{(\bigcup_i w_i)}(s)                  &==& \SI{i \in I}\ISA{w_i}(s)  &\quad {= s |= \ISA{(\bigcup_i w_i )}}\\
        \ISD{(\bigcup_i w_i)}\big(s,<i,a>\big)    &==& \ISD{w_i}(s,a)            \\
        \ISn{(\bigcup_i w_i)}\big(s,<i,a>,d\big)  &==& \ISn{w_i}(s,a,d)
      \ee
      and
      \be
      \ISA{(\bigcap_i w_i)}(s)                  &==& \PI{i \in I}\ISA{w_i}(s) &\quad {=  s |= (\bigcap_i \ISA{w_i})}\\
        \ISD{(\bigcap_i w_i)}(s,f)                &==& \SI{i \in I}\ISD{w_i}(s,a)\\
        \ISn{(\bigcap_i w_i)}\big(s,f,<i,d>\big)  &==& \ISn{w_i}\big(s,f(i),d\big) \ \hbox{.}
      \ee

  \item[Composition]
      Suppose $w_1$ and $w_2$ are interaction structures $S_1 -> \Fam^2(S_2)$ and
      $S_2 -> \Fam^2(S_3)$; define a structure $w_1\SEQ w_2$, 
       called the \emph{sequential composition} of $w_1$ and $w_2$. having type
      $S_1->\Fam^2(S_3)$ with components:
      \be
       A(s_1)                           &==&  \SI{a_1\in A_1(s_1)}                                      \\
                                        &  &  \PI{d_1\in D_1(s_1,a_1)} A_2(s_1[a_1/d_1])                \\
       D\big(s_1,<a_1,f>\big)           &==&  \SI{ d_1\in D_1(s_1,a_1)}D_2\big(s[a_1/d_1],f(d_1)\big)   \\
       n\big(s_1,<a_1,f>,<d_1,d_2>\big) &==&  s_1[a_1/d_1][f(d_1)/d_2]                                  \ \hbox{.}
      \ee
      \ie~a command in $\ISA{(w_1\SEQ w_2)}(s)$ is given by a command in
      $\ISA{w_1}(s)$, and a continuation $f$ giving, for all responses $d$ in
      $\ISD{w_1}(s,a)$ a command in $\ISA{w_2}(s[a/d])$.  
Note that $\ISA{(w_1 \SEQ w_2)} = {w_1}^\circ(\ISA{w_2})$. 

    \item[Unit]
      \be
        \ISA{\SKIP}(s)               &==& \One  \\
        \ISD{\SKIP}(s,\BLANK)        &==& \One  \\
        \ISn{\SKIP}(s,\BLANK,\BLANK) &==& s     \ \hbox{.}
      \ee
    \end{description}
\end{defn}
These operations satisfy the expected laws:
\begin{prop} \label{prop:ISPT}
  \be
    \SKIP^\circ              &=& \SKIP                        \ \hbox{;}\\
    \angelU{T}^\circ         &=& \angelU{T^\circ}             \ \hbox{;}\\
    \demonU{T}^\circ         &=& \demonU{T^\circ}             \ \hbox{;}\\
    (\bigcup_i w_i)^\circ    &=& \bigcup_i ({w_i}^\circ)      \ \hbox{;}\\
    (\bigcap_i w_i)^\circ    &=& \bigcap_i ({w_i}^\circ)      \ \hbox{;}\\
    (w_1 \SEQ w_2)^\circ     &=& {w_1}^\circ \SEQ {w_2}^\circ \ \hbox{;}\\
    (w^\circ)^\bot           &=& \comp · w^\circ · \comp   & \hbox{\textbf{(only classically)}} \ \hbox{.} \ee
\end{prop}
\begin{proof}
  Routine.  Note that though to define the relation $T^\circ$ requires use of 
  equality, one can define the predicate transformers $\angelU{T^\circ}$ and
  $\demonU{T^\circ}$ without it.

  \noindent
  For the last point, we have constructively that
  \[
    s |= (w^\circ)^\bot(V) \quad\hbox{iff}\quad (\forall U \sub S)\ s |=
    w^\circ(U) => U\olp V
  \]
  which can be taken as the definition of the dual for an arbitrary monotonic
  predicate transformer. This variant is better behaved in a constructive
  setting, and classically equivalent to the $\comp·\BLANK·\comp$ definition.

\end{proof}

\smallbreak
In view of this proposition, we may regard interaction structures as concrete
representations of monotone predicate transformers that support many useful
operators of the refinement calculus. (Iteration will be dealt with in
subsection \ref{sec:ISIterations} \vpageref{sec:ISIterations}.) As a result,
we allow ourselves to overload the name $w$ of an interaction structure to
mean also $w^\circ$.

\subsubsection{Factorization of interaction structures}

It is worth observing that any interaction structure $w: S -> \Fam^2(S'')$ is equal
to the composition $\angelU{T_a} \SEQ \demonU{T_d}$ where $T_a : S
-> \Fam(S')$, $ T_d : S' -> \Fam(S'') $ and $ S' = \SI{ s \in S} \ISA{w}(s)$.
The transition structures $T_a$ (which ``issues the command'') and $T_d$
(which ``performs the command'') are defined as follows:
\be
  \TSI{T_a}         &==& \ISA{w}  &\qquad& \TSI{T_d}\big(<s,a>\big)     &==& \ISD{w}(s,a) \\
  \TSn{T_a}(s,a)    &==& <s,a>    &      & \TSn{T_d}\big(<s,a>,d\big)   &==& \ISn{w}( s, a, d)
\ee
This factorization should be compared with the normal form theorem for
predicate transformers mentioned \vpageref{normalFormTh}.  Just as $\angelU{R_a}
\SEQ \demonU{R_d}$ is a normal form for monotone predicate transformers, so
(with transition structures replacing relations) it is a normal form for
interaction structures.

In this connection, one can define a symmetric variant of the notion of
interaction structure, consisting of two arbitrary sets $S$ and $S'$ with either
(i) a pair of relations between them, or (ii) a pair of transition structures in
opposite directions.
We have used the name ``Janus structure'' for type-(ii) structures
(based on transition structures in different directions).  Markus
Michelbrink has used the name ``interactive game'' for type-(i) structures. 
Michelbrink's work shows that these to be highly interesting
structures.  The relation they bear to monotone predicate
transformers seems not unlike that the natural numbers bear to the (signed) integers.

\subsection{Iteration} \label{sec:ISIterations}

We now define the iterative constructs $\RTC\BLANK$ and $\FISH\BLANK$ on
interaction structures.

\subsubsection{Angelic iteration}

\begin{defn}\label{defn:angelic-iteration}
  Let $w : S -> \Fam^2(S)$; define
  \smallbreak
  \noindent
  $\begin{array}{lll}
    \ISA{\RTC{w}} &==&
      \MU{X : S -> \Set}\quad \LAM{s \in S}\\
        & & \DATA{\Exit \\
                  \Call(a,f) \WHERE{  a \in S(s) \\
                                      f \in \PI{d \in D(s,a)} X(s[a/d]) }}
   \end{array}$
  \smallbreak
  \noindent
  $\begin{array}{lll}
       \ISD{\RTC{w}}(s,\Exit)              &==& \DATA{\Nil} \\
       \ISD{\RTC{w}}\big(s,\Call(a,f)\big) &==& \DATA{ \Cons(d,d')}
                                                  \WHERE{ d \in D(s,a) \\
                                                          d' \in \RTC{D}\big(s[a/d],f(d)\big)}\\
   \end{array}$
  \smallbreak
  \noindent
  $\begin{array}{lll}
       \ISn{\RTC{w}}(s,\Exit,\Nil)  &==&\  s \\
       \ISn{\RTC{w}}\big(s,\Call(a,f),\Cons(d_0,d')\big) &==&\ \ISn{\RTC{w}}\big(s[a/d_0],f(d_0),d'\big)
   \end{array}$
\end{defn}

\noindent
An element of $\RTC{A}(s)$ is a data-structure that can be interpreted as a
program or strategy for the Angel, to issue commands and react to the Demon's
responses to commands. Because the definition uses a least fixpoint, this
program is well founded in the sense that the Angel eventually reaches an
$\Exit$ command.

Associated with each such program $p \in \RTC{A}(s)$, the set $\RTC{D}(s,p)$
and the function $\RTC{n}(s,p,\BLANK)$ give the family of states in which it
can exit.  Elements of the former can be seen as paths from $s$ through $p$,
while the latter maps a path to its final state.  An element of $\RTC{D}(s,p)$
is sometimes called a (finite and complete) run, log, trace, or history.  Note
that a trace is intrinsically finite.

\begin{prop}  \label{prop:ISAngelIteration}
  For any interaction structure $w$ on $S$, we have
  $ {\RTC{w}}^\circ = \RTC{{w^\circ}} $.
\end{prop}
\begin{proof}
  Easy inductive proof.
\end{proof}

To make formulas easier to read, we adopt Sambin's ``$<|$'' notation:
\begin{defn}
  If $w:S->\Fam^2(S)$, $s\in S$, and $U,V\sub S$, put:
  \be
    s <|_w U &==& s|={\RTC{w}}^\circ(U) \ \hbox{;} \\
    V <|_w U &==& V \sub {\RTC{w}}^\circ(U)\ \hbox{.}
  \ee
\end{defn}
This higher-order relation satisfies:
\begin{lem} \leavevmode
  \begin{enumerate}
    \item monotonicity: $s <|_w U \ \hbox{and}\ U \sub V => s <|_w V$;
    \item reflexivity: $s |= U => s <|_w U$;
    \item transitivity: $s <|_w U \ \hbox{and}\ U<|_w V => s <|_w V$.
  \end{enumerate}
\end{lem}
\begin{proof}
  This is a just a rewriting of the definition of a closure operator using the
  ``$<|$'' notation. (${\RTC{w}}^\circ$ is a closure operator by
  proposition~\ref{prop:FISH_RTC_int_cl}, since
  ${\RTC{w}}^\circ=\RTC{{w^\circ}}$.)
  \\
  Note that this proof (that ${\RTC{w}}^\circ$ is a closure operator) is entirely
  predicative.

\end{proof}

\subsubsection{Demonic iteration}

We first recall  the rules used in~\cite{setzerhancock:venice2003} to generate
``state dependent'' greatest fixpoints. Translated to our setting, if
$<A,D,n>$ is an interaction structure on~$S$, we are allowed to form the
family $\FISH{A}$ of sets indexed by $s\in S$ using the following rules:
\begin{itemize}

  \item formation rule: \RULE{s\in S}{\FISH{A}(s) : \Set};

  \item introduction rule: (setting up a coalgebra)\\
    \RULE{X:S -> \Set & F:X \sub w^\circ(X) & s\in S & x\in X(s)}%
         {\Coiter(X,F,s,x) \in \FISH{A}(s)};\\
    \COMMENT{(recall that $F:X\sub w^\circ(X)$ means $F:\PI{s} X(s) -> \SI{a}\PI{d}X(s[a/d])$)}

  \item elimination rule:
    \RULE{s\in S & K\in\FISH{A}(s)}%
         {\Elim(s,K) \in w^\circ(\FISH{A})(s)};

  \item reduction rule: 
    \[
      \Elim\big(s,\Coiter(X,F,s,x)\big)
         = \begin{array}[t]{l}
           \left(a,\LAM{d}\Coiter\big(X,F,s[a/d],g(d)\big)\right) \\
           \WHERE{<a,g> = F(s,x)}       \ \hbox{.}
           \end{array}
    \]
  (Here ``$<a,k> = \ldots$'' is how we indicate an implicit pattern matching.)

\end{itemize}

It should be noted (\cite[p. 11]{setzerhancock:venice2003}) that those
rules (which require that a weakly final coalgebra for $w^\circ$) are
dual to the rules for inductive types.  Roughly speaking, they are the
coinductive analogue of Petersson and Synek's inductively defined
treeset constructions, expressed with a specific destructor
$\Elim$.\footnote{ Implicit in these rules is a certain ``weak''
  impredicative existential quantifier, that permits the formation of
  the higher product $\SI{X:\Set} A : \Set$, but without the strong
  projections of the usual Sigma type. Instead, one has an elimination
  rule closer to that in traditional natural deduction.  Such a
  ``weak'' quantifier is sometimes invoked in the analysis of abstract
  data types (\cite{MitchellPlotkin88}).}

\begin{defn}\label{defn:demonic-iteration}
  Let $w : S -> \Fam^2(S)$; define
  \smallbreak
  \noindent
  $\begin{array}{lll}
    \ISA{\FISH{w}} &==&
       \NU{X : S -> \Set} \quad \LAM{ s\in S}\\
       & & \SI{a\in\ISA{w}(s)} \PI{d\in\ISD{w}(s,a)} X(s[a/d])\\
       &==& \FISH{A}
   \end{array}$

  \smallbreak
  \noindent
  $\begin{array}{lll}
       \ISD{\FISH{w}} &==&
          \MU{ X : \PI{s \in S} \FISH{A}(s) -> \Set } \quad\LAM{ s \in S, p \in \FISH{A}(s) }\\
          & &
             \DATA{ \Nil \\
                    \Cons(d,d') \quad
                         \WHERE{
                                  <a,k> = \Elim(p) \\
                                  d \in D(s,a) \\
                                  d' \in X(s[a/d],k(d))
                               }}
   \end{array}$

  \smallbreak
  \noindent
  $\begin{array}{lll}
       \ISn{\FISH{w}}(s,p,\Nil)                &==& s \\
       \ISn{\FISH{w}}\big(s,p,\Cons(d,d')\big) &==& \ISn{\FISH{w}}\big(s[a/d],k(d),d'\big)\\
                                               &  & \quad \WHERE{<a,k> = \Elim(p)}
   \end{array}$
\end{defn}

An element of $\FISH{A}(s)$ can be interpreted as a command-response program
starting in state $s$ and continuing for as many cycles as desired, perhaps
forever.  One can picture such a program as an infinite tree, in which control
flows along a branch in the tree.  An element of $\FISH{D}(s,p)$ is a finite
sequence of responses that may be returned to the agent running the program
$p$; and $\FISH{n}(s,p,t)$ is the state obtained after the finite response
sequence $t$ has been processed.

\begin{prop} \label{prop:ISDemonIteration}
  For any interaction structure $w$ on $S$, we have
  ${\FISH{w}}^\circ = \FISH{{w^\circ}}$.
\end{prop}
\begin{proof}
  Let $U\sub S$:
  \begin{itemize}

    \item  ${\FISH{w}}^\circ(U) \sub \FISH{{w^\circ}}(U)$: since
      $\FISH{{w^\circ}}(U)$ is the greatest fixpoint of $U\cap
      w^\circ(\BLANK)$, it suffices to show that ${\FISH{w}}^\circ(U)$ is a
      post-fixpoint for the same operator, \ie~ that ${\FISH{w}}^\circ(U)
      \sub U \cap w^\circ\big({\FISH{w}}^\circ(U)\big)$.
      \\
      Let $s|={\FISH{w}}^\circ(U)$; this implies that there is some $p\in
      \FISH{A}(s)$ \st
      \[
        \big(\forall t\in\FISH{D}(s,p)\big)\ s[p/t] |= U \ \hbox{.}
      \]
      In particular, for $t=\Nil$, we have $s[p/\Nil]=s|=U$.

      We now show that $s |= w^\circ\big({\FISH{w}}^\circ(U)\big)$.  Let
      $\Elim(p)$ be of the form $<a_0,k>$. We claim that $\big(\forall d\in
      D(s,a_0)\big)\ s[a_0/d]|={\FISH{w}}^\circ(U)$:
      if $d\in D(s,a_0)$, we have $k(d)\in \FISH{A}(s[a_0/d])$ and
      $\Cons(d,d')\in\FISH{D}\big(s,<a_0,k>\big)$ for any $d'$ in
      $\FISH{D}\big(s[a_0/d],k(d)\big)$.  This implies (because
      $s|={\FISH{w}}^\circ(U)$) that
      \[
        s[a_0/d][k(d)/d'] \quad=\quad s[<a_0,k>/\Cons(d,d')]\quad|=\quad U
      \]
      which completes the proof.

    \item $\FISH{{w^\circ}}(U) \sub {\FISH{w}}^\circ(U)$: let
      $s|=\FISH{{w^\circ}}(U)$;\\
      we need to find a $p\in\FISH{A}(s)$
      \st~$\big(\forall t\in\FISH{D}(s,p)\big)\ s[p/t]|=U$.  By the
      introduction rule for $\FISH{A}$, it suffices to find a coalgebra
      $(X:S->\Set\,,\,F)$ with $F \in X \sub w^\circ X$.
      \\
      $X==\FISH{{w^\circ}}(U)$ together with the function $F$ coming from the
      coinductive rule $\FISH{{w^\circ}} \sub \SKIP\cap(w^\circ\SEQ
      \FISH{{w^\circ}}) \sub w^\circ\SEQ \FISH{{w^\circ}}$ is such a coalgebra.
      \\
      This provides us with an element
      $
        \Coiter\big(X,F,s,x\big) \in\FISH{A}(s)
      $
      where $x$ is the proof that $s|=\FISH{{w^\circ}}(U)$.

      We will show the following: ``for all states $s$, for all programs $p$
      generated by this coalgebra, for all responses $t$ to $p$, we have
      $s[p/t]|=U$''.  More precisely, we will prove:
      \[
        \Big(\forall s\,,\
             \forall x\in X(s) \,,\
             \forall t\in\FISH{D}\big(s,p(s,x)\big)\Big)
        \quad s[p(s,x)/t] |= U
      \]
      where $p(s,x)=\Coiter(X,F,s,x)$.
      \\
      We work by induction on the structure of $t$.
      \begin{description}

        \item[base case:] if $t=\Nil$, then $s[p(s,x)/\Nil] = s$, and we have
          $s|=U$ since $s|=X=\FISH{{w^\circ}}(U)\sub U$.

        \item[induction case:] if $t=<d_0,t'>$, then $s[p(s,x)/<d_0,t'>] =
          s[a_0/d_0][k(d_0)/t']$ where $\Elim\big(p(s,x)\big)=<a_0,k>$. By the
          reduction rule for coinduction, we have; if $x$ is of the form
          $<a_0,f>$:
          \[
            \Elim\,\Coiter(X,F,s,x) = \big(a_0,\LAM{d_0}
            \Coiter\big(X,F,s[a_0/d_0],f(d_0)\big)\big)
          \]
          Therefore, $k(d_0)=p\big(s[a_0/d_0],f(d_0)\big)$, and we obtain the
          result by applying the induction hypothesis for $s[a_0/d_0]$,
          $f(d_0)\in X(s)$ and
          $t'\in\FISH{D}\big(s,p(s[a_0/d_0],f(d_0))\big)$.

      \end{description}
  \end{itemize}
\end{proof}

\begin{cor}
  For any interaction structure  $w$, we have ${\FISH{{w^\bot}}}^\circ =
  \FISH{{w^\bullet}}$.
\end{cor}
\begin{proof}
 Direct from lemma~\ref{lem:ISDual} and proposition~\ref{prop:ISDemonIteration}.
\end{proof}

Just as for $\RTC{w}$ and $<|$, we introduce the following notation:
\begin{defn}
  If $w$ is an interaction structure on $S$, put:
  \be
    s |><_w U &==& s|={w^\bot}^\infty(U) \ \hbox{;} \\
    V |><_w U &==& V \olp {w^\bot}^\infty(U) \ \hbox{.}
  \ee
\end{defn}

\subsection{Clients, servers and their interaction} \label{sec:clientServer}

In the vast majority of cases, there are only two kinds of program one is
called upon to write: in programming terminology, those are called client
programs and server programs. For background, see \cite{Schneider90}.
Clients and servers are agents on opposite sides of a service interface,
sometimes also called a resource interface.  The service may be, for example,
to store values in addressable memory cells, or disk sectors. The client
obtains or uses the service, the server provides it.  In general terms, the
behavior of an agent following a client program is to issue commands across
the interface, and then use the responses to steer control to the right
continuation point in the program, iterating through some finite number of
command-response cycles until eventually reaching an exit point in the
program.  On the other hand, the behavior of an agent following a server
program is to wait passively for a command, perform it and respond
appropriately, for as many command-response cycles as required by the client.

The programming terminology of ``clients'' and ``servers'' is connected with
the angelic and demonic forms of iteration described above in section
\ref{sec:ISIterations}.   The client issues requests
or commands, and the server performs them and responds to the client with a
sequence of results, one for each issued command.  Each request, its performance,
and the response to it constitutes a \emph{command-response cycle.} From the
client's perspective, we may think of the performance of the request as an
atomic event that occurs sometime between issuing the request and receiving
the response.  The server changes state, as it were ``in a trice''.

\smallbreak
A server may have many clients.  As when someone is operating a till in a
supermarket, we may arrange (or simulate in various ways) that a client has
the exclusive attention of a server, cycling through the purchase of several
items by a single client, until the trolley is empty, the customer pays, and
an entire \emph{transaction}, consisting of many cycles is complete.  Then the
next customer in the queue comes forward.  The number of cycles is at the
discretion of the client.  In essence what is happening here is that the
server performs an entire transaction program (whose execution consists of
several cycles) which we can view as a single composite command. The response
to this composite command is a record or trace of responses to the individual
commands: as it were, the receipt handed to the supermarket customer when the
transaction is complete.  However, what is important is that the transactions
\emph{appear} to take place in a total order.  Outside of supermarkets, there
are ways of processing transactions such that several transactions can be in
progress, and their commitment is scheduled to optimize either throughput or
response time.  Essentially, starting a transaction is not something visible,
and one can always pretend that transactions are started the instant before
they are committed.

To describe clients and servers only in such a mechanistic way is
however to miss something important. A client or server program is
written to accomplish some purpose, or to fulfill an intention.  The
purpose or intention is expressed by a specification, ideally a formal
specification that can be handled by a machine and used in
verification.  The crucial question is: what are the logical forms of
the specifications of client and server programs?  The interest of
dependent type theory as a framework for developing programs is that
one may hope, by exploiting the expressive power of the type system,
to express specifications formally and yet with full precision.  One
may then harness decidable type-checking to guide the development of
programs to meet those specifications.

\medbreak
Let us attempt to answer this question.  What follows is merely an attempt to
summarize experience of reading and writing specifications for both client and
server programs.

\smallbreak
Suppose $w$ describes an interface; a \newterm{client program} is specified by
a pair:
\begin{description}
  \item[$\Init \sub S$:]
    a predicate describing initial states in which the program is required to
    work. (In other states the program need not even terminate.) The user of
    the program is obliged to ensure that the initial predicate holds before
    running the program.

  \item[$\Next \sub S × S$:]
    a relation defined between initial states and final states.  The value of
    $\Next$ for states outside $\Init$ is irrelevant: the behavior of the
    program is unspecified. Very often (but not always) this relation has the
    simpler ``rectangular'' form $\Init × \Goal$ for some $\Goal \sub S$;
    meaning that the goal predicate does not depend on the initial state.

\end{description}

A client program satisfying such a specification is in essence a constructive
proof that $ \Init \sub \SET{ s \in S }{ s <|_w \Next(s) } $.  When the
$\Next$ relation happens to be of the form $\Init × \Goal$, this takes the
simpler form
\[ \Init \quad <|_w\quad \Goal\ \hbox{.} \]

If we have such a proof, and the interface is in a state $s$ such that
initial predicate $\Init$ holds, then we can use the proof as a guide
or strategy to bring about a state in which the goal predicate
$\Next(s)$ holds, if only we are provided with a server that responds
to all our requests.

\bigbreak
As for server programs, the situation is the following: again, let $w$
describe the interface.  A \newterm{server program} is usually described by a
pair of predicates:
\begin{description}

\item[$\Init \sub S$:]
  a non-empty set which describes the allowed initial states of the service.

\item[$\Inv \sub S$:]
  a predicate that holds initially and is maintained by the server.

\end{description}

\begin{remark}
  By symmetry with specifications $\Init \sub \SET{s \in S}{s <|_w \Next(s)
  }$, where the relation $\Next$ is not necessarily rectangular, one may also
  consider server specifications of the form $\Init \olp \SET{s \in S}{s |><_w
  \Next(s) }$.  At first sight the general case seems to have no counterpart
  in practice.  However, if $\Next$ is actually a simulation relation one can
  express a certain kind of recoverability with a specification of this more
  general form.  (This is connected with the discussion of localization at
  \vpageref{sec:localdist}.)
\end{remark}

A program satisfying such a specification is in essence a constructive proof
that $\Init$ overlaps with the weakest post-fixpoint (invariant) of $w^\bot$ included in $\Inv$.
That is to say, it yields a state
together with a proof that the state belongs to both the initial predicate and
that invariant. 
Recall lemma~\ref{lem:ISDual} that if $w$ is
given by an interaction structure
\[
  U \sub w^\bot(U)
  \quad<=>\quad
  (\forall s|=U)\big(\forall a\in A(s)\big)\  \big(\exists d\in D(s,a)\big)\ s[a/d]|=U
  \ \hbox{.}
\]
In other words the Demon is never deadlocked, but can always respond to any
legal command, and moreover in such a way that the invariant continues to hold in
the new state.

Note that a direct consequence of lemma~\ref{lem:invFISH_satRTC} is that any
invariant can be written in the form $\FISH{(w^\bot)}(V)$.  The predicate $V$
need not itself be an invariant, but can be weaker than the actual
invariant $\FISH{(w^\bot)}(V)$, and so \textit{a fortiori} is maintained by
the server program. 

To summarise, a server specification takes the form $\Init
\olp \FISH{{w^\bot}}(\Inv)$ where $\Inv$ is a predicate guaranteed to hold
before and after every step.
Using the $|><$ notation, this gives:
\[
  \Init |><_w \Inv \ \hbox{.}
\]

\paragraph{Interaction between client and server programs.}

What happens when we put a client and a server program together, and run the
former ``on'' the latter?  The answer is connected with the compatibility rule
in Sambin's formalization of basic topology.

Suppose that in some state $s$ of a common interface $w$, we have a client
program $P$ that can be run to bring about a goal predicate $U$ (\ie~$s<|U$),
and a server program $K$ that maintains a predicate $V$ (\ie~$s|><V$).  When
all internal calculation has been carried out, the client program $P$ will
have been brought into one of two forms: either $<\Call(a,f),g>$ where
\be
    a \in A(s) \\
    f \in \PI{d \in D(s,a)} \RTC{A}(s[a/d]) \\
    g \in \PI{<d_0,d'> \in \RTC{D}\big(s,\Call(a,f)\big)}\ s[a/d][f(d)/d']|=U
    \ \hbox{,}
\ee
or $<\Exit,h>$ where $h(\Exit)$ is a proof that $s |= U$.  On the other hand,
if $<K,l>$ is the \emph{server} program, then $\Elim(s,K)$ has the form
$<r,k>$ where
\be
      r \in \PI{a \in A(s)} D(s,a)   \\
      k \in \PI{a \in A(s)} (A^\bot)^\infty(s[a/r(a)]) \\
      l \in \PI{t \in (D^\bot)^\infty(s,K)} s[K/t] |= V
      \ \hbox{.}
\ee
For any $U,V\sub S$, we define an \newterm{execution} function
 with the type
 \[
  \exec_{U,V}\big(
  <s,P,K> |= \RTC{w}(U) \olp {w^\bot}^\infty(V)
  \big)
  \quad \in \quad
  U \olp {w^\bot}^\infty(V)
\]
by means of the following clauses:
\be
  \exec_{U,V}\big(s, \, <\Exit,h>      ,\, <K,l> \big) &==& <s,h(\Exit),K> \\
  \noalign{\smallbreak}
  \exec_{U,V}\big(s, \, <\Call(a,f),g> ,\, <K,l> \big) &==& \\
  \noalign{\smallbreak}
  \noalign{\hskip3.5cm
         \begin{array}[t]{ll}
           \textbf{let} &\begin{array}[t]{lcl}
                               <r,k> &=& \Elim(s,K)\\
                               d  &==& r(a) \\
                               P' &==& f(d)  \\
                               g' &==& \LAM{d'} g(<d,d'>)\\
                               K' &==& k(a)  \\
                               l' &==& \LAM{t} l(<a,t>)\\
                            \end{array} \\
                            \textbf{in} &\exec_{U,V}\big(s[a/d],\, <P',g'> ,\, <K',l'>\big)
                          \end{array}}
\ee
If we strip away the parameters and programs from this rule, we obtain
\begin{center} \label{execution}
  \RULE{\RTC{w}(U) \olp \FISH{{w^\bot}}(V)}{U \olp \FISH{{w^\bot}}(V)}
\end{center}
that can immediately be recognized as Sambin's compatibility rule
(\cite{BPIII}).  In some sense this rule expresses the mechanics of
interaction between client and server programs.

\medbreak
How does this rule apply to the formulas given above for the general form of
client and server specifications? Suppose we have a client program satisfying
the specification ``$\Init <|_w \Goal$'', and a server program satisfying the
specification ``$\Init |><_w \Inv$''.  Then we can apply the execution function
to get:
\begin{center}
  \RULE{\Init <| \Goal  &  \Init |>< \Inv}%
       {\Goal |>< \Inv} \ \hbox{.}
\end{center}
The real use of a client program is turn servers in a state that 
satisfies the precondition into servers in a state
that satisfies the goal predicate. 

\paragraph{Safety and Liveness.}

The concepts of partial and total correctness emerged from the investigations
of Floyd, Dijkstra and Hoare 
into the foundations of specification and
verification for sequential programming.  A program is partially correct if it
terminates only when it has attained the goal that it should, while it is totally
correct if in addition it terminates whenever it should.  In the late 70's,
Lamport in \cite{Lamport:1977:PCM} introduced the terms \newterm{safety} and
\newterm{liveness} as the appropriate generalizations of these concepts to the
field of concurrent programming.  In concurrent programming a program
interacts with its environment while it is running, rather than only when
initialized or terminated.  Informally, a safety property requires that
``nothing bad'' should occur during execution of a concurrent program. (A time
can be associated with the violation of a safety property).  On the other hand
a liveness property requires that ``something good'' should occur (so that it
is violated only at the end of time, as it were).
These properties soon received formal definitions, in the case of safety by
Lamport \cite{lamport85:_formal_found_specif_verif}, and in the case of
liveness by Alpern and Schneider \cite{alpern85:_defin_liven}.

\smallbreak
These properties were defined in topological terms, with respect to the
``Baire'' space of infinite sequences of states. (The set of sequences sharing
a common finite prefix is a basic neighborhood in this space).
%
Briefly, a safety property was analyzed as a closed set of sequences,
and a liveness property as a dense set (\ie~one intersecting with
every non-empty open set).  The properties were also expressed in
terms of linear-time temporal logic, the idea being that a safety
property asserts that something is (now and) \emph{forever} the case,
whereas a liveness property requires that something (now or)
\emph{eventually} takes place.  For various reasons liveness is
usually restricted to fairness properties in which the temporal
modalities are nested at most twice. 
An example of a fairness requirement is so-called ``strong''
fairness, which requires that an event (state-change) of a certain kind occurs
infinitely often providing that it is enabled infinitely often.  A
readable account of the r\^ole these concepts play in practical
specification can be found in Lamport's book
\cite{lamport02:spec_systems}.


\smallbreak
What can we say about these notions from the perspective of interaction
structures?  One thing that can be said with some confidence is that a safety
property is an invariant.  In basic topology, invariants represent closed
sets.  So this agrees with Lamport's topological analysis.

A liveness property on the other hand is merely a set of points which overlaps with
every non-empty open set.  It seems difficult to say anything interesting
about liveness properties in general; but it may be easier when the
properties are simple combinations of particular modalities such as ``infinitely
often'' and ``eventually always''.

\subsection{Product operations} \label{sec:products}

We describe below two product operations on interaction structures.  The first
corresponds to an operation treated in the refinement calculus
(\cite{productRC}), while the second does not.

\paragraph{Synchronous tensor.}

Suppose $w_1$ and $w_2$ are two interaction structures on $S_1$ and $S_2$.
We define $w_1 \otimes w_2$ on $S_1×S_2$:  \label{def:synchTensor}
\be
  \ISA{(w_1 \otimes w_2)}(<s_1,s_2>)
                    &==&  \ISA{w_1}(s_1) \times \ISA{w_2}(s_2)
                    \\
  \ISD{(w_1 \otimes w_2)}(<s_1,s_2>,<a_1,a_2>)
                    &==&  \ISD{w_1}(s_1,a_1) \times  \ISD{w_2}(s_2,a_2)
                    \\
  \ISn{(w_1 \otimes w_2)}(<s_1,s_2>,<a_1,a_2>,<d_1,d_2>)
                    &==&  <s_1[a_1/d_1],s_2[a_2/d_2]>
\ee
The computational meaning of this operation is clear: one issues commands in
each of a pair of interfaces, receives responses from them both, and they each
move to their new state, simultaneously and atomically. Sometimes this kind of
arrangement is called ``ganging'', or ``lock-step synchronization''.

The synchronous tensor corresponds to the following operation on predicate
transformers (addition to propositions~\ref{prop:ISPT},
\ref{prop:ISAngelIteration} and~\ref{prop:ISDemonIteration}):
\[
    (F_1 \otimes F_2)(R) = \bigcup_{U × V \sub R} F_1(U) × F_2(V)
\]
which was used in~\cite{productRC} to model parallel execution of program
components.

In combination with duality (definition~\ref{def:ISDual}), the synchronous
tensor enjoys strong algebraic properties
(see~\cite{PTlinear}).

\paragraph{Angelic product.}

Similarly, suppose $w_1$ and $w_2$ are two interaction structures on $S_1$ and
$S_2$.  We define $w_1 \odot w_2$ on $S_1×S_2$:
\be
  \ISA{(w_1 \odot w_2)}(<s_1,s_2>)               &==&  \ISA{w_1}(s_1) + \ISA{w_2}(s_2) \\
  \ISD{(w_1 \odot w_2)}(<s_1,s_2>,\INL{a_1})     &==& \ISD{w_1}(s_1,a_1)               \\
  \ISD{(w_1 \odot w_2)}(<s_1,s_2>,\INR{a_2})     &==& \ISD{w_2}(s_2,a_2)               \\
  \ISn{(w_1 \odot w_2)}(<s_1,s_2>,\INL{a_1},d_1) &==& <s_1[a_1/d_1],s_2>               \\
  \ISn{(w_1 \odot w_2)}(<s_1,s_2>,\INR{a_2},d_2) &==& <s_1,s_2[a_2/d_2]>               \ \hbox{.}
\ee
The computational meaning is again quite clear: a pair of interfaces is
available to the Angel, who choose the one to use.  This kind of
arrangement is frequently found at the low-level interface of a program
component, where instances of various resources are exploited, one at a time,
to implement a higher-level interface.  We call this kind of combination the
``angelic product''.

In terms of predicate transformers, the angelic product corresponds to
\[
    (F_1 \odot F_2)(R) = \bigcup_{\SING{s_1} × V \sub R} \SING{s_1} × F_2(V)
                         \quad \cup
                         \bigcup_{U × \SING{s_2} \sub R} F_1(U) × \SING{s_2} \quad.
\]



\section{Morphisms}
\label{sec:morphisms}

\subsection{Linear simulations}

We now consider what to take for morphisms between predicate
transformers or their representation by interaction structures.  The
definition we adopt coincides with what is known as a ``forward''
simulation in the refinement calculus.  As we will see in section
\ref{sec:topology}, it is also connected with the definition of
continuous relation in formal topology.

\smallbreak
Let us therefore consider the case of (homogeneous) interaction structures,
subscripted with ``$h$'' and ``$l$'' to distinguish the high and low
level interfaces.
\[
  \begin{array}{ccccc}
    w_h &:& S_h &->& \Fam^2(S_h) \\
       \downarrow            \\
    w_l &:& S_l &->& \Fam^2(S_l)
  \end{array}
\]
As explained earlier, we view $w_h$ and $w_l$ as command-response interfaces
over the state spaces $S_h$ and $S_l$, where the command and response
``dialects'' are given by $<A_h,D_h>$ and $<A_l,D_l>$ respectively.  Our
intuition here is to think of a morphism as a systematic translation between
the dialect for $w_h$ and the dialect for $w_l$, which enables us to use a
device supporting the interface $<S_l,w_l>$ as if it were a device supporting
the interface $<S_h,w_h>$.  That is, we should be able to translate high level
$A_h$-commands into low level $A_l$-commands, and responses to the latter (low
level $D_l$ responses) back into high level $D_h$ responses in such a way that
the simulation of $<S_h,w_h>$ by $<S_l,w_l>$ can be indefinitely sustained.

\smallbreak
It is often the case that several different low-level states can represent the
same high-level state, so that the link between high-level states and
low-level states can be represented by a function from the latter to the
former (sometimes called an abstraction function, or refinement mapping).  It
is also sometimes the case that several different high-level states can be
represented by the same low-level state.  For such reasons, many people take
the link between high and low level states to be a general relation, rather
than a map one one direction or the other.

\smallbreak
The question then is: how can we make this intuition of translation precise?
The answer we propose is the following.

\begin{defn}
  Let $w_h : S_h -> \Fam^2(S_h)$, and $w_l : S_l -> \Fam^2(S_l)$.  A
  \newterm{linear simulation} of $<S_h,w_h>$ by $<S_l,w_l>$ is a relation $R
  \sub S_h×S_l$ which satisfies the following ``sustainability'' condition:

\begin{tabbing}
  xx\=xx\=xxxxxxxxxxxxxxxxxxxxxxxxx\= \kill
  If \> $(s_h,s_l)|=R$, then  \\
     \>$\forall a_h \in A_h(s_h)$          \>\>\small-- for all high-level commands $a_h$ \ldots  \\
     \>$\exists a_l \in A_l(s_l)$          \>\>\small-- there is a low-level command $a_l$ s.t.\\
     \>$\forall d_l \in D_l(s_l,a_l)$      \>\>\small-- for all responses $d_l$ to the low-level command \ldots \\
     \>$\exists d_h \in D_h(s_h,a_h)$      \>\>\small-- there is a response $d_h$ to the command $a_h$ \st\\
     \>\>$\big(s_h[a_h/d_h],s_l[a_l/d_l]\big)|=R$  \>  \small-- the simulation can be sustained.
\end{tabbing}
We write $R:w_h -o w_l$ to mean $R$ is a linear simulation from $w_h$ to $w_l$.
\end{defn}

In explanation of the qualification ``linear'', we have required a one-for-one
intertranslation between the high and low-level interfaces.  (We shall shortly
introduce a notion of general simulation, allowing zero or non-zero low-level
interactions for each high-level interaction.)

The formula above with its four nested quantifiers is perhaps a little
daunting at first sight.  Let's re-express it in a more compact form.

\begin{lem} \label{lem:linSimAlternative}
  $R \sub S_h×S_l$ is a linear simulation of $<S_h,w_h>$ by $<S_l,w_l>$ iff
  for all $s_h \in S_h$, and $a_h \in A_h(s_h)$, we have $R(s_h) \sub
  w_l\big(\bigcup_{d_h \in D_h(s_h,a_h)} R(s_h[a_h/d_h])\big)$.
\end{lem}
\begin{proof}
  Simple formal manipulation.
\end{proof}

\begin{remark}
  A linear simulation from $w_h$ to $w_l$ is itself an invariant for a certain
  relation transformer ``$w_h -o w_l$''.

  \begin{defn}
    If $w_h$ and $w_l$ are interaction structures on $S_h$ and $S_l$, define a
    new interaction structure on $S_h×S_l$ with:
  \be
    A(<s_h,s_l>)             &==& \SI{f \in A_h(s_h) -> A_l(s_l)}                        \\
                             &  & \PI{a_h \in A_h(s_h)}\ D_l(s_l,f(a_h)) -> D_h(s_h,a_h) \\
    D(<s_h,s_l>,<f,g>)       &==& \SI{a_h \in A_h(s_h) }\ D_l(s_l,f(a_h))                \\
    n(<s_h,s_l>,<f,g>,<a_h,d_l>) &==&  <s_h[a_h/g(a_h,d_l)],s_l[f(a_h)/d_l]>             \ \hbox{.}
  \ee
  \end{defn}
  \noindent
  This concrete representation is merely the result of applying the axiom of
  choice to pull the quantifier alternation $\PI{\BLANK}
  \SI{\BLANK}\PI{\BLANK} \SI{\BLANK}$ into $\SI{\BLANK}\PI{\BLANK}$ form.
  Notice that everything has a computational meaning: the commands are
  intertranslation functions, the responses are data outside the control of
  the simulation, and data is communicated between the high and low poles of a
  state-pair.
  \smallbreak
  \noindent
  Classically, this interaction structure is (isomorphic to) the
  representation of the linear-logic implication from~\cite{PTlinear}. The
  corresponding tensor is the synchronous tensor~$\otimes$ defined
  \vpageref{def:synchTensor}. (One can check that ``$\otimes$'' is
  left-adjoint to ``$-o$''.) It is interesting to remark that neither
  composition nor iteration of predicate transformers/interaction structures
  are used in the models of linear logic from \cite{PTlinear,ISlinear}.

\end{remark}

\medbreak
The following proposition gives a characterization of linear simulations as a
subcommutativity property (point~\textit{\ref{prop:linSimChar:point:subcom}}).
\begin{prop} \label{prop:linSimChar}
  The following are equivalent:
  \begin{enumerate}
    \item $R$ is a linear simulation of $<S_h,w_h>$ by $<S_l,w_l>$;
    \item\label{prop:linSimChar:point:subcom} $\angelU{\CONV{R}} \SEQ
      w_h  \sub w_l \SEQ \angelU{\CONV{R}}$;
    \item\label{prop:linSimChar:point:last} for all $U\sub S_h$, $s_h\in S_h$ we have
       $s_h<|_{w_h} U => R(s_h) <|_{w_l} R(U)$.
\end{enumerate}
\end{prop}
\begin{proof}
  \COMMENT{(the implication 2\/$=>$1 requires the use of equality)}

  \textit{1}$=>$\textit{2}:
  we have to show that $s_l |= (R \SEQ w_h)(U)$ implies $s_l |=
  (w_l \SEQ R)(U)$.
  \smallbreak
  \begin{algproof}
    $s_l |= R \SEQ w_h(U)$

    \step{<=>}{definition of $\SEQ$}

    $(\exists s_h\in S_h)\ (s_h,s_l)|= R \ \hbox{and}\ s_h |= w_h(U)$

    \step{=>}{definition of the predicate transformer $w_h$}

    $(\exists s_h)\ %
      \begin{array}[t]{l}
        (s_h,s_l)|= R \ \hbox{and} \\
        \big(\exists a_h\in A_h(s_h)\big)\,\big(\forall d_h\in D_h(s_h,a_h)\big)\ s_h[a_h/d_h]|=U
      \end{array}$

    \step{=>}{by lemma~\ref{lem:linSimAlternative}}

    $(\exists s_h)\ %
      \begin{array}[t]{l}
        (s_h,s_l)|= R \ \hbox{and} \\
        s_l |= w_l\big(\bigcup_{d_h} R(s_h[a_h/d_h])\big) \ \hbox{and}\ \bigcup_{d_h} s_h[a_h/d_h] \sub U
    \end{array}$

    \step{=>}{$R=\angelU{\CONV{R}}$ commutes with unions}

    $(\exists s_h)\ %
      \begin{array}[t]{l}
        (s_h,s_l)|= R \ \hbox{and} \\
        s_l |= w_l R\big(\bigcup_{d_h} s_h[a_h/d_h]\big) \ \hbox{and}\ \bigcup_{d_h} s_h[a_h/d_h] \sub U
      \end{array}$

    \step{=>}{by monotonicity}

    $s_l |= w_l\SEQ R(U)$
  \end{algproof}

  \medbreak
  \textit{2}$=>$\textit{1}:
  suppose that $R\SEQ w_h \sub w_l\SEQ R$, and let $s_h\in S_h$ and $a_h\in
  A_h(s_h)$; we will show that $R(s_h) \sub w_l\big(\bigcup_{d_h \in
  D_h(s_h,a_h)} R(s_h[a_h/d_h])\big)$ and conclude using
  lemma~\ref{lem:linSimAlternative}.
  \\
  Define $U==\bigcup_{d_h\in D_h(s_h,a_h)}\SING{s_h[a_h/d_h]}$. (This where
  equality is needed.)
  \\
  We certainly have that $s_h|=w_h(U)$ so that $R(s_h)\sub R\SEQ w_h(U)$. By
  hypothesis, this implies that $R(s_h)\sub w_l\SEQ R(U)$ which we had set out
  to prove. (Since $R$ commutes with unions.)

  \smallbreak
  The proof that \textit{2}$<=>$\textit{3} is straightforward.

\end{proof}
The following is easy:
\begin{prop}  \label{prop:simClosedUnion}
  If $w_1$ and $w_2$ are interaction structures, the linear simulations of
  $w_1$ by $w_2$ are closed under arbitrary unions (including the empty union,
  so that there is always an empty simulation).
\end{prop}

Finally, the following shows that we have a poset enriched category.
\begin{prop}\label{prop:simCategory}\leavevmode
  \begin{enumerate}
    \item The relational composition $(R_1 \SEQ R_2)$ of two linear
      simulations is a linear simulation.
    \item If $w$ is an interaction structure on $S$, then $\eq_S:w -o w$.
    \item Composition of linear simulations is monotone in both its arguments.
  \end{enumerate}
  We call this category $\LinSim$.
\end{prop}
\begin{proof}
  Straightforward.
\end{proof}
\noindent
The same proposition holds if we replace interaction structures with predicate
transformers, and use point~\textit{\ref{prop:linSimChar:point:subcom}} from
proposition~\ref{prop:linSimChar} as the definition of simulation. We call
this category $\PT$.

\begin{remark}
  Of course to define a category, we need equality relations for the identity
  morphisms of this category.  Without equality, we have a weaker structure,
  having merely an associative and monotone composition of morphisms.
\end{remark}

A morphism is supposed to ``preserve structure''.  What is the structure
preserved by a simulation?  The following observation suggests one answer.
\begin{lem}
  If $R$ is a simulation as above, the image of an
  invariant for ${w_h}$ is an invariant for ${w_l}$, \ie~the image
  of a high-level invariant is a low-level invariant.
\end{lem}
\begin{proof}
  simple application of proposition~\ref{prop:linSimChar}.
\end{proof}

\begin{remark}
  The notion of a linear simulation is already well-known in the literature of
  the refinement calculus (see for example
  \cite{back00:_encod_decod_data_refin}). There it is known as forward (or
  ``downward'') data refinement.  In fact, in that setting one considers a
  more general notion, in which the relation (which may be identified with a
  disjunctive predicate transformer) is generalized to a ``right-moving''
  predicate transformer:
   \begin{defn}
     If $F_h$ and $F_l$ are transformers, and if $P :\Pow(S_l)
    ->\Pow(S_h)$, then  $F_h$ is said to be data-refined through $P$ by $F_l$
    if
    \[
      P \SEQ F_h   \sub   F_l \SEQ P \ \hbox{.}
    \]
    If $P$ commutes with arbitrary unions, then the refinement is said to be
    ``forward'', whereas if $P$ commutes with arbitrary intersections, the
    refinement is said to be ``backward''.
  \end{defn}

  In the setting of impredicative higher-order logic one can prove that the
  predicate transformers that commute with arbitrary unions are precisely
  those of the form $\angelU{Q}$ for some relation $Q$, and those that commute
  with arbitrary intersections are precisely those of the form $\demonU{Q}$.
  It follows that a linear simulation is a forward data-refinement.  It is
  natural to wonder whether one can give a predicative analysis of backward
  data refinement, akin to that we have given of forward data refinement.
\end{remark}


\subsection{Monads and general simulations}
\label{subsec:general-simulations}

When a high-level interface is implemented on top of a low-level, less
abstract interface, it is rare that a single high-level command (for example:
record this data as a file in such and such a directory) can be translated to
a \emph{single} low-level command. Instead, several interactions across the
low-level interface (reading and writing disk sectors) are usually required
before the high-level operation can be completed.  In essence, what we are
going to do is make the notion of simulation more flexible and applicable by
moving to the Kleisli category for a certain monad.

\smallbreak
There are at least three monads of interest: the reflexive closure, the
transitive closure and the reflexive/transitive closure.

\begin{description}

\item[RC]
  The functor $\RC(F) = \SKIP \cup F$ is monadic.  A morphism in the Kleisli category
  from $(S_1,F_1)$ to $(S_2,F_2)$ is a linear simulation of $(S_1,F_1)$ by
  $\big(S_2,\RC(F_2)\big)$, which we call an \newterm{affine simulation} of
  $(S_1,F_1)$ by $(S_2,F_2)$.  A step in $(S_1,F_1)$ need not make use of a
  step in $(S_2,F_2)$.

\item[RTC]
  $\RTC{\BLANK}$ is monadic.
  A morphism in the Kleisli category from $(S_h,F_h)$ to $(S_l,F_l)$ is a
  linear simulation of $(S_h,F_h)$ by $(S_l,\RTC{F_l})$, which we call a
  \newterm{general simulation} of $(S_h,F_h)$ by $(S_l,F_l)$.  A step in
  $(S_h,F_h)$ may make use of any number of steps in $(S_l,F_l)$.

\item[TC]
  The functor $\TC{F} = F \SEQ \RTC{F}$ is monadic.  A morphism in its Kleisli
  category is a linear simulation of $(S_h,F_h)$ by $\big(S_l,\TC{F_l}\big)$.
  It translates high-level commands to low level programs that run for at least one step.

\end{description}

\begin{prop}
  $\RC(\BLANK)$, $\TC{\BLANK}$ and $\RTC{\BLANK}$ are monads in $\LinSim$ and
  $\PT$. We call the Kleisli category of $\RTC{\BLANK}$ the category of
  general simulations and interaction structures: $\GenSim$. We write $R:w_h
  -> w_l$ for morphisms in this category (\ie~$w_h->w_l$ is a synonym for $w_h
  -o \RTC{w_l}$).
\end{prop}
\begin{proof}
  We will work with interaction structures; the case of predicate transformer
  is very similar. Moreover, we only treat the case of the $\RTC{\BLANK}$
  functor; the other cases being similar.

  \smallbreak
  Recall that an endofunctor $M$ on a category $\mathcal{C}$ is a monad (in
  triple form) if we have the following:
  \begin{itemize}
    \item an operation $\BLANK^\sharp$ taking any
      $f:\mathcal{C}\big[A,M(B)\big]$ to an
      $f^\sharp:\mathcal{C}\big[M(A),M(B)\big]$;
    \item for any object $A$ of $\mathcal{C}$, a morphism
      $\eta_A:\mathcal{C}\big[A,M(A)\big]$
  \end{itemize}
  such that:
  \begin{enumerate}
    \item $\eta\SEQ f^\sharp = f$;
    \item ${\eta_A}^\sharp = \Id_{M(A)}$;
    \item $(f\SEQ g^\sharp)^\sharp = f^\sharp\SEQ g^\sharp$.
  \end{enumerate}
  It is trivial to check that $\eq:w -o \RTC{w}$; and the next proposition
  will show that if $R$ is a linear simulation of $w_h$ by $\RTC{w_l}$ then
  $R$ is a linear simulation of $\RTC{w_h}$ by $\RTC{w_l}$. Thus we can put:
  $R^\sharp == R$ and $\eta_{<S,w>} == \eq_S$.

\end{proof}

\begin{lem} \label{lem:genSimChar}
  Let $w_h$ and $w_l$ be two interaction structures on the sets $S_h$
  and~$S_l$; let $R$ be a relation on $S_h×S_l$. The following are equivalent:
  \begin{enumerate}
    \item $R$ is a linear simulation $w_h -o \RTC{w_l}$;
    \item for any $s_h\in S_h$ and $a_h\in A_h(s_h)$:
      \[R(s_h)  <|_{w_l}  \bigcup_{d_h\in D_h(s_h,a_h)} R\big(s_h[a_h/d_h]\big)\ \hbox{;}\]
    \item \label{lem:genSimChar:pt:RTC}
      for any $s_h\in S_h$ and $a'_h\in \RTC{A_h}(s_h)$:
      \[R(s_h)  <|_{w_l}  \bigcup_{d'_h\in \RTC{D_h}(s_h,a'_h)} R\big(s_h[a'_h/d'_h]\big)\ \hbox{.}\]
  \end{enumerate}
\end{lem}
\begin{proof}
  In turn:
  \smallbreak

    \textit{1}$<=>$\textit{2}:
    simple consequence of proposition~\ref{prop:linSimChar}.

    \smallbreak
    \textit{3}$=>$\textit{2}:
    follows from the observation that $\eq_{S_h}$ is a linear simulation $w_h
    -o \RTC{w_h}$.

    \smallbreak
    \textit{2}$=>$\textit{3}:
    let $s_h\in S_h$ and $a'_h\in \RTC{A_h}(s_h)$; we do the proof by
    induction on $a'_h$:
    \begin{description}
      \item[base case]
        if $a'_h = \Exit$, then we only need to show that $R(s_h) <|_{w_l}
        R(s_h)$, which is trivially true since $\SKIP \sub \RTC{w_l}$.

      \item[induction case]
        if $a'_h$ is of the form $\Call(a_h,f_h)$, then we have:
        \begin{itemize}
          \item $R(s_h)  <|  \bigcup_{d_h} R\big(s_h[a_h/d_h]\big)$;
            \COMMENT{(by point 1$<=>$2 of this lemma)}

          \item for any $d_h\in D_h(s_h,a_h)$, by induction hypothesis, we have:
            \[
              R\big(s_h[a_h/d_h]\big)
              \quad<|_{w_l}\quad
              \bigcup_{d'_h}
                R\big(s_h[a_h/d_h][f_h(d_h)/d'_h]\big)
            \]
            \COMMENT{where $d'_h \in \RTC{D_h}(s_h[a_h/d_h],f_h(d_h))$}

          \item since the RHS is a subset of $\bigcup_{d_h,d'_h}
            R\big(s_h[\Call(a_h,f_h)/<d_h,d'_h>]\big)$ we can conclude (by
            monotonicity) that
            \[
              R\big(s_h[a_h/d_h]\big)
              \quad<|_{w_l}\quad
              \bigcup_{d'_h} R\big(s_h[a'_h/d'_h]\big)
            \]
            which, by transitivity, implies
            \[
               R(s_h)
               \quad<|_{w_l}\quad
               \bigcup_{d'_h} R\big(s_h[a'_h/d'_h]\big)
            \]
        \end{itemize}
    \end{description}
\end{proof}

\begin{cor}
  We have: $R$ is a linear simulation $w_h -o \RTC{w_l}$ iff $R$ is a linear
  simulation $\RTC{w_h} -o \RTC{w_l}$.
\end{cor}

\subsection{Saturation, equality of morphisms}
\label{sec:saturation}

We have argued that the category~$\GenSim$ serves as a model for component
based programming.  However, the notion of equality on morphisms is still too
strong.  It may be that two general simulations differ extensionally though
they still have the same potential, or ``simulative power''.
\begin{defn}
  If $R$ is a general simulation from $w_h$ to $w_l$, we define the following
  relation $\Sat{R}$ (the saturation of $R$) on $S_h×S_l$:
  \[
  (s_h,s_l) |= \Sat{R}  \quad==\quad  s_l |= \RTC{w_l}·R(s_h)
  \ \hbox{.}
  \]
\end{defn}
\noindent
This amounts to considering instead of functions $R:S_h -> \Pow(S_l)$,
functions $R: S_h -> \mathit{Sat}(w_l)$, where $\mathit{Sat}(w_l)$ is the
collection of $w_l$-saturated predicates. (See
lemma~\ref{lem:invFISH_satRTC}.)

The intuition behind saturation is the following. Suppose $R$ is a relation
between low level states $S_l$ and high level states $S_h$. The saturation of
$R$ is a relation which allows ``internal'' or ``hidden'' low level
interaction.  To simulate a high level state $s_h$ by a low level state $s_l$,
it is permissible that the Angel has a program that constrains interactions
starting in $s_l$ to terminate in states that simulate $s_h$.

\medbreak
\noindent
We have:
\begin{prop} \label{prop:sat}
  Let $R$ be a general simulation of $w_h$ by $w_l$, then $\Sat{R}$ is
  also a general simulation of $w_h$ by $w_l$.
\end{prop}

\begin{proof}
  According to lemma~\ref{lem:genSimChar}, we need to show $\Sat{R}(s_h) <| \bigcup_{d_h}
  \Sat{R}\big(s_h[a_h/d_h]\big)$.

  \smallbreak
  By lemma~\ref{lem:genSimChar}, we have $R(s_h) <| \bigcup_{d_h}R\big(s_h[a_h/d_h]\big)$
  and since ${\RTC{w_2}}$ is a closure operator, we have
  \[
    \RTC{w_2}\big(R(s_h)\big) \equiv \Sat{R}(s_h)
    \quad <| \quad
    \bigcup_{d_h}R\big(s_h[a_h/d_h]\big)\ \hbox{.}
  \]
  For any $d_h$, $R\big(s_h[a_h/d_h]\big) <| R\big(s_h[a_h/d_h]\big)$
  which implies  (still because $\RTC{w_2}$ is a closure operator)
  \[
    R\big(s_h[a_h/d_h]\big)
    \quad <| \quad
    \RTC{w_2}\Big(R\big(s_h[a_h/d_h]\big)\Big) \equiv \Sat{R}(s_h[a_h/d_h])
    \ \hbox{.}
  \]
  Since the above is true for any $d_h$, it implies that
  \[
    \bigcup_{d_h} R\big(s_h[a_h/d_h]\big)
    \quad <| \quad
    \bigcup_{d_h}\Sat{R}\big(s_h[a_h/d_h]\big)
    \ \hbox{.}
  \]
  We get the result by transitivity.

\end{proof}

Thus, saturation provides us with an appropriate ``normalization'' operation
when comparing general simulations: to compare two simulations, compare their
normal forms.  So we put:

\begin{defn}  \label{def:saturation}
  Let $R_1$, $R_2$ be two general simulations of $w_h$ by $w_l$; we say that
  \begin{itemize}
    \item $R_2$ is \newterm{stronger} than $R_1$ (written $R_1 \sqsub
      R_2$) if $\Sat{R_1} \sub \Sat{R_2}$;
    \item $R_1$ is \newterm{equivalent} to $R_2$ ($R_1 \approx R_2$) if $R_1
      \sqsub R_2$ and $R_2 \sqsub R_1$.
  \end{itemize}
\end{defn}
\noindent
The following is trivial: (point~\textit{\ref{pt:satClosure}\/} follows from
proposition~\ref{prop:sat}).
\begin{lem}
  We have:
  \begin{enumerate}
    \item $\sqsub$ is a preorder on the collection of general simulations from
      $w_h$ to $w_l$;
    \item $\approx$ is an equivalence relation;
    \item\label{pt:satClosure} $\Sat{R}$ is (extensionally) the largest
      relation in the equivalence class of $R$;
    \item the operation $R |-> \Sat{R}$ is a closure operation.
  \end{enumerate}
\end{lem}

\noindent
We can now conclude this section:
\begin{prop}
  $(\GenSim,\sqsub)$ is a poset enriched category.
\end{prop}
\begin{proof}
  The only thing we need to check is that composition is monotonic in both its
  arguments.\footnote{This result, together with all the required lemmas has
  been formally checked using the Agda system.}

  \medbreak

  Let $R_1$, $R_2$ be two simulations of $w_h$ by $w_m$ and $Q_1$, $Q_2$ two
  simulations of $w_m$ by $w_l$ such that $R_1\sqsub R_2$ and $Q_1\sqsub Q_2$.
  Suppose moreover that $s_h \in S_h$; we need to show that $\Sat{R_1\SEQ
  Q_1}(s_h) \sub \Sat{R_2\SEQ Q_2}(s_h)$:

  \begin{itemize}

  \item we have $\Sat{R_1}(s_h) \sub \Sat{R_2}(s_h)$ because $R_1 \sqsub R_2$;

  \item we also have $\big(Q_1\SEQ \Sat{R_1}\big)(s_h) <|_l
    \big(Q_2\SEQ\Sat{R_2}\big)(s_h)$:

  let $s_l |= \big(Q_1\SEQ\Sat{R_1}\big)(s_h)$, \ie~$(s_m,s_l)|=Q_1$ for some
  $s_m$ s.t. $(s_h,s_m)|=\Sat{R_1}$. We will show that $s_l <|_l
  \big(Q_2\SEQ\Sat{R_2}\big)(s_h)$:
    \begin{itemize}
      \item $Q_2(s_m) \sub \big(Q_2\SEQ\Sat{R_2}\big)(s_h)$ since $s_m |= \Sat{R_1}(s_h)
        \sub \Sat{R_2}(s_h)$;
      \item $s_l |= \Sat{Q_2}(s_m)$ because $Q_1\sqsub Q_2$ and
        $s_l|=\Sat{Q_1}(s_m)$; \COMMENT{(since $s_l|= Q_1(s_m)$)}
      \item so by monotonicity, $s_l |= \big(\RTC{w_l}\SEQ
        Q_2\SEQ\Sat{R_2}\big)(s_h)$.
    \end{itemize}

  \item From the last point, we get $\big(\RTC{w_l}\SEQ Q_1\SEQ\Sat{R_1}\big)(s_h) \sub
    \big(\RTC{w_l}\SEQ Q_2\SEQ\Sat{R_2}\big)(s_h)$;

  \item for any simulation $R:w -o \RTC{w'}$, we have
    $\big(\RTC{w'}\SEQ R\SEQ\RTC{w}\big)(U) = \big(\RTC{w'}\SEQ R\big)(U)$:
    \begin{itemize}
      \item[{\footnotesize$\sub$}:] because $\big(R\SEQ\RTC{w}\big)(U) \sub \big(\RTC{w'}\SEQ
        R\big)(U)$ and
        $\RTC{w'}$ is a closure operator;
      \item[{\footnotesize$\supseteq$}:] $\SKIP \sub \RTC{w} => \RTC{w'}\SEQ R
        \sub \RTC{w'}\SEQ R \SEQ \RTC{w}$.
    \end{itemize}

  \item So we can conclude:
    \[
      \Sat{R_1\SEQ Q_1}(s_h)
      \equiv
      \big(\RTC{w_l}\SEQ Q_1\SEQ R_1\big)(s_h)
      \sub
      \big(\RTC{w_l}\SEQ Q_2\SEQ R_2\big)(s_h)
      \equiv \Sat{R_2\SEQ Q_2}(s_h)
      \ \hbox{.}
    \]
  \end{itemize}
\end{proof}

%


\section{The link with formal topology}\label{sec:topology}

Our title mentions both programming and formal topology.  We now (at last)
turn to the topological meaning of our constructions. We start by recalling
the most basic notions of formal topology.

\subsection{Formal and basic topology}

The aim of formal topology was to develop \newterm{pointfree topology} in a
fully constructive (\ie~predicative) setting. Motivations for pointfree
topology can be found in~\cite{Pointless}. Briefly, pointfree topology studies
the properties of the lattice of open sets of a topology, without ever
mentioning points (hence the name).  Many traditional topological theorems are
classically equivalent to a pointfree version that can be proved
constructively without the axiom of choice. Example of such theorems include
Hahn-Banach theorem, Heine-Borel theorem, or various representation theorems
(such as Stone's).  The idea is thus to factor out all non-constructive
methods into the proof that the pointfree version is equivalent to the
traditional theorem.

\smallbreak
\newterm{Basic topology} amounts to removing the condition of distributivity
of the lattice of open sets. The result is a very concise and elegant
structure which, surprisingly enough, still contains the basic notions of
topology (closed sets, open sets and continuity).  It is the basis of a
modular approach to formal topology in that one can add exactly what is
needed in order to understand a particular property.

Introductions to the subject can be found in
\cite{Pointless,topologyvialogic,firstcommunication,BP,SP,BPIV}.

\subsubsection{Basic topologies}

See \cite{BPII} for details.
\smallbreak
\noindent
\begin{defn}
  A \newterm{basic topology} is a set $S$ together with two predicate
  transformers $\A$ and $\J$ on $S$ such that:
\begin{itemize}
  \item $\A$ is a closure operator;
  \item $\J$ is an interior operator;
  \item $\A$ and $\J$ are \emph{compatible}:
    \smash{\RULE{\A(U) \olp
    \J(V)}{U\olp\J(V)}} for all $U,V \sub S$.
  \medskip
\end{itemize}
\end{defn}

The set $S$ is intended to represent a base of the topology; and so, an
element $s\in S$ will be called a \newterm{formal basic open.} A subset $U$ of
$S$ is called \newterm{open} when $U=\A(U)$; and a subset $V$ of $S$ is called
\newterm{closed} when $V=\J(V)$.\footnote{No mistakes: a formal open is
\emph{closed} in the sense of $\A$; and a formal closed is open in the sense
of $\J$! See \cite{BPI} for the justification.}

\medbreak
A minimal requirement is that open sets [resp. closed sets] form a sup lattice
[resp. inf lattice]. This is indeed the case:

\begin{lem} \label{lem:openLattice}
  If $(U_i)_{i\in I}$ is family of open sets, define $\bigvee_{i}
  U_i=\A\big(\bigcup_{i\in I} U_i\big)$; the type of open sets with
  $\bigvee$ and $\cap$ is a lattice with all set-indexed sups.
  \smallbreak
  \noindent
  If $(V_i)_{i\in I}$ is family of closed sets, define $\bigwedge_{i}
  V_i=\J\big(\bigcap_{i\in I} V_i\big)$; the type of closed sets with
  $\bigwedge$ and $\cup$ is a lattice with all set-indexed infs.
\end{lem}
However, these lattices are generally speaking not distributive.  (We will see
a way to add distributivity in section~\ref{sec:distributivity}.)  As a consequence
there is no notion of \emph{point} in basic topology.\footnote{More
precisely, without distributivity the notion of a point cannot be
distinguished from that of a closed subset!}

\subsubsection{Formal continuity}

See \cite{BPIV} for details.
\smallbreak
\noindent
Since a continuous function from $(S_1,\A_1,\J_1)$ to $(S_2,\A_2,\J_2)$ should
map open sets in $(S_2,\A_2,\J_2)$ to open sets in $(S_1,\A_1,\J_1)$, it
cannot be represented directly by a function from $S_2$ to $S_1$.  A
continuous function has to be represented by a relation between $S_1$ and
$S_2$. If $f \sub S_1×S_2$ represents such a continuous function, the
intuitive, concrete meaning of $(s_1,s_2)|=f$ is thus ``$s_1 \sub
f^{-1}(s_2)$'', where $s_1$ and $s_2$ are basic opens.

\begin{defn} \label{def:continuity}
  If $(S_{1},\A_{1},\J_{1})$ and $(S_{2},\A_{2},\J_{2})$ are basic topologies,
  and $R$ a relation between $S_{1}$ and $S_{2}$; $R$ is \emph{continuous} if
  the two conditions hold:
  \begin{enumerate}
    \item $\CONV{R}\big(\A_2(V)\big) \sub \A_1\big(\CONV{R}(V)\big)$;
    \item $R\big(\J_1(U)\big) \sub \J_2\big(R(U)\big)$.
  \end{enumerate}
\end{defn}
\noindent
Equivalent characterizations are listed in~\cite{BPIV}. It is worth
noting that the two conditions are in general independent.

\medbreak
By definition, two continuous relations $R$ and $T$ from $S_1$ to $S_2$ are
(topologically) equal if $\A(\CONV{R}s_2) = \A(\CONV{T}s_2)$ for all $s_2\in
S_2$.  The main purpose of this definition is to remove dependency on the
specific ``base'' of the topology considered.

Basic topologies and continuous relations with topological equality form a
category which is called~$\BFTop$.

\subsubsection{Convergent basic topologies} \label{sec:distributivity}

See \cite{BPIII} for details.
\smallbreak
\noindent
The above structure still lacks many properties found in ``real'' topologies;
in particular, the binary infimum need not distribute over arbitrary suprema.
One way to get distributivity is to add the following condition on the
operator~$\A$:

\begin{defn} \label{def:convergence}
  Let $\A$ be a closure operator on a set $S$; write $U\BinDown V$ for the
  subset $\big\{s \  | \  (\exists s'|= U)\ s |= \A\{s'\} \hbox{  and  }
  (\exists s''|= V)\ s |= \A\{s''\}\big\}$. We say that $\A$ is
  \newterm{convergent} if the following holds:
\begin{center} \label{convergence}
  \RULE{s|=\A(U) & s|=\A(V)}%
       {s|=\A(U\BinDown V)}
  \ \hbox{.}
\end{center}
\end{defn}
\noindent
This condition is sometimes called \newterm{summability of approximations:} it
gives a way to compute the intersection of two open sets from their
representatives. If $U$ and $V$ represent the two open sets $\A(U)$ and
$\A(V)$,\footnote{It is a trivial observation that a subset is open iff it is
of the form $\A(U)$.} then $U\BinDown V$ represents the intersection
$\A(U)\cap\A(V)$.

\begin{lem} \label{lem:convDistr}
  If $(S,\A,\J)$ is a convergent basic topology, then its lattice of open sets
  is distributive.
\end{lem}
\begin{proof}
  For any $U\sub S$, define $\downclosure{U}=\big\{s\in S\ |\ (\exists s'|=
  U)\,s|=\A\{s'\}\big\}$. We have $U\BinDown V=\downclosure{U} \cap
  \downclosure{V}$.


  \smallbreak
  Let $U$ be an open set and $(V_{i})_{i\in I}$ a set-indexed family of open
  sets; \ie~we have $U=\A(U)$ and $V_i=\A(V_i)$ for all $i\in I$.

  \smallbreak
  \begin{algproof}
    $\bigvee_{i\in I} U \cap V_i$

    \step{=}{$U$ and the $V_i$'s are open}

    $\bigvee_{i\in I} \A(U)\cap\A(V_i)$

    \step{=}{convergence}

    $\bigvee_{i\in I} \A(U\BinDown V_i)$

    \step{=}{definition of $\bigvee$ and easy lemma: $\A\bigcup\A = \A\bigcup$}

    $\A\Big(\bigcup_{i\in I} U\BinDown V_i\Big)$

    \step{=}{distributivity of $\cap$ and $\bigcup$}

    $\A\Big(\downclosure{U} \cap \bigcup_{i\in I} \big(\downclosure{V_i}\big)\Big)$

    \step{=}{}

    $\A\Big(U\BinDown \big(\bigcup_{i\in I} V_i\big)\Big)$

    \step{=}{convergence}

    $\A(U) \cap \A\Big(\bigcup_{i\in I} V_i\Big)$

    \step{=}{a union of open sets is an open set}

    $U \cap \bigvee_{i\in I} V_i$
  \end{algproof}
  \smallbreak
  \noindent
  which completes the proof that open sets do indeed form a frame.

\end{proof}
Traditionally, formal topologies are also equipped with a \newterm{positivity
predicate} called~$\Pos$.  Its intuitive meaning is ``$s|=\Pos$
iff $s$ is non-empty''.
This predicate was required to satisfy the \newterm{positivity axiom}:
\COMMENT{(where $U^+ = U \cap \Pos$)}
\begin{center}
  \RULE{s|=\A(U^+)}{s|=\A(U)}
\end{center}
which means that only positive opens really contribute to the topology.

The positivity predicate is now defined from $\J$: $\Pos==\J(S)$, and the
positivity axiom is not required anymore. (Though it will hold in all examples
with a real topological flavor.)

\medbreak
In a convergent basic topology, we can define the notion of point:
(see~\cite{BPIV})
\begin{defn}
  Let $(S,\A,\J)$ be a convergent basic topology; a subset $\alpha \sub S$ is
  said to be a~\newterm{point} if:
  \begin{enumerate}
    \item $\alpha$ is closed: $\alpha = \J(\alpha)$;
    \item $\alpha$ is non-empty: $\alpha\olp\alpha$;
    \item $\alpha$ is convergent: $s_1|=\alpha\,,\,s_2|=\alpha =>
      \SING{s_1}\BinDown \SING{s_2} \olp \alpha$.
  \end{enumerate}
\end{defn}

\subsection{The topology of an interaction structure}

Interaction structures can be viewed as an ``interactive'' reading of the
notion of \emph{inductively generated topology}.

\subsubsection{Basic topology}

Recall that if $w=<A,D,n>$ is an interaction structure on $S$,
propositions~\ref{prop:ISAngelIteration}, \ref{prop:ISDemonIteration} and
\ref{prop:FISH_RTC_int_cl} guarantee that:
\begin{itemize}
  \item $\RTC{w}$ is a closure operator on the subsets of $S$;
  \item $\FISH{{w^\bot}}$ is an interior operator on the subsets of $S$.
\end{itemize}
We also have the execution formula (page~\pageref{execution}):
  \RULE{\RTC{w}(U) \olp \FISH{{w^\bot}}(V)}{U \olp \FISH{{w^\bot}}(V)}.

\noindent
As a result, we put:
\begin{defn}
  If $w$ is an interaction structure on $S$, define:
  \be
    \A_w(U) &==& \RTC{w}(U)\ \hbox{;}\\
    \J_w(U) &==& \FISH{{w^\bot}}(U)\ \hbox{.}
  \ee
\end{defn}
\noindent
We have:
\begin{lem}
If $w:S->\Fam^2(S)$, then $(S,\A_w,\J_w)$ is a basic topology.
\end{lem}

\bigbreak
In \cite{induc_top}, the authors use the notion of axiom set to inductively
generate a formal cover. The difference between axiom sets and interaction
structures is merely that an axiom set is an element of the type
$S->\Fam\big(\Pow(S)\big)$ that was mentioned
\vpageref{discussionAlternativeIS}.

If we look at the rules used to generate $<|_w$, \ie~for $\RTC{w}$, we obtain:
\begin{itemize}
  \item\infer{s|= U}%
             {s|=\A(U)}%
             {\small $\Exit$};
  \item\infer{a\in A(s) & \big(\forall d\in D(s,a)\big) \, \big(n(s,a,d)|=\A(U)\big)}%
             {s|=\A(U)}%
             {\small $\Call$}.
\end{itemize}

Those correspond exactly to the \textit{reflexivity} and \textit{infinity}
rules used in \cite{induc_top} to generate the cover ``$<|$''.

\subsubsection{Continuous relations revisited}

We argued above that (generated) basic topologies and interaction structures
are the same notions with different intuitions.  We will now lift the notion
of continuity to the realm of interaction structures.  The result is that in
basic topology, continuous relations are exactly general simulations
(proposition~\ref{prop:sim_continuous} and lemma~\ref{lem:eqTopSat}).

\smallbreak
\noindent
Before anything else, let's prove a little lemma about the $\J$ operator:
\begin{lem} \label{lem:GenSimSubComJ}
  Suppose $w_h$ and $w_l$ are interaction structures, and $R$ is a general
  simulation of $w_h -> w_l$. Then $\CONV{R}·\J_l(V) \sub \J_h·\CONV{R}(V)$
  for all $V\sub S_l$.
\end{lem}

\begin{proof}
  Suppose that $V\sub S_l$, $(s_h,s_l)|=R$ and $s_l|=\J_l(V)$; we need to show
  that $s_h|=\J_h\big(\CONV{R}(V)\big)$. Since $\J_h\big(\CONV{R}(V)\big)$ is
  the greatest fixpoint of the operator $\big(\CONV{R}(V)\big) \cap
  w^\bot(\BLANK)$, it suffices to show that $s_h$ is in a pre-fixpoint of
  the same operator. We claim that $\CONV{R}(V)$ is such a pre-fixpoint:

  \begin{itemize}

    \item $s_h |= \CONV{R}(V)$ because $(s_h,s_l)|=R$ and $s_l|=\J_l(V)\sub
      V$;
    \item $\CONV{R}(V) \sub \CONV{R}(V)$;
    \item $\CONV{R}(V) \sub w^\bot\big(\CONV{R}(V)\big)$:

      \noindent
      let $s_h\in\CONV{R}(V)$ and $a_h\in A_h(s_h)$; we need to find a $d_h\in
      D_h(s_h,a_h)$ \st~$s_h[a_h/d_h]|=\CONV{R}(V)$.\\
      By lemma~\ref{lem:genSimChar}, we know that
      $s_l|=\A_l\big(\bigcup_{d_h}R(s_h[a_h/d_h])\big)$, and because
      $s_l|=\J_l(V)$, we can apply the execution formula to obtain a final
      state $s'_l|= \bigcup_{d_h}R(s_h[a_h/d_h]) \cap \J_l(V)$.  In
      particular, there is a $d_h\in D_h(s_h,a_h)$ such that $s_l'|=
      R(s_h[a_h/d_h])$.\\
      Since $s'_l|=\J_l(V)\sub V$, it implies that $s_h[a_h/d_h]|=\CONV{R}(V)$.
  \end{itemize}
\end{proof}

\noindent
With this new lemma, it is easy to prove the following:
\begin{prop} \label{prop:sim_continuous}
  Let $w_h$ and $w_l$ be two interaction structures, let $R$ be a relation
  between $S_h$ and $S_l$; $R$ is a general simulation $w_h -> w_l$ iff
  $\CONV{R}$ is a continuous relation from $(S_l,\A_l,\J_l)$ to
  $(S_h,\A_h,\J_h)$.
\end{prop}
\begin{proof}
Suppose first that $\CONV{R}$ is continuous; the definition implies in
particular that $R\big(\A_h(U)\big)\sub \A_l\big(R(U)\big)$ for all $U\sub
S_h$. By lemma~\ref{lem:genSimChar}, this implies that $R$ is a general
simulation from $w_h$ to $w_l$.

\smallbreak\noindent
The converse is a direct application of lemma~\ref{lem:genSimChar} and
lemma~\ref{lem:GenSimSubComJ}.
\end{proof}

\medbreak
The category of basic topologies and continuous relation $\BFTop$ has a notion
of equality which is more subtle (though coarser) than plain extensional
equality of relations.  Transposing it in our context we get: $R\approx Q$ if
and only if $\A\big(R(s_h)\big)=\A\big(Q(s_h)\big)$ for all $s_h\in S_h$.

\begin{lem} \label{lem:eqTopSat}
  If $R$ and $Q$ are simulations, then $R$ and $Q$ are topologically equal iff
  their saturations are extensionally equal. ($R$ and $Q$ have the same
  potential, see page~\pageref{def:saturation}.)
\end{lem}

\subsubsection{Topological product}

In section~\ref{sec:products}, we introduced a notion of binary ``angelic
tensor'', morally corresponding to the union of several interaction
structures.  This operation was already defined in~\cite{induc_top} (and
probably in other places) as the product topology.  In particular, we have the
two continuous projection relations.
\begin{lem} \label{lem:AtensorProj}
  If $w_1$ and $w_2$ are two interaction structures, then the two following
  relations
  \begin{itemize}
    \item $\pi_1 = \{(s_1,(s_1,s_2)) \in S_1×(S_1×S_2)\}$;
    \item $\pi_2 = \{(s_2,(s_1,s_2)) \in S_2×(S_1×S_2)\}$
  \end{itemize}
  are (linear) simulations from $w_i$ to $w_1\odot w_2$ (for $i=1,2$); and
  \textit{a fortiori} are morphisms in all the categories considered.
\end{lem}
If one interprets the sets $S_1$ and $S_2$ as (pre)bases, and $<|$ are the
covering relation; it is clear that this corresponds indeed to the usual
product of topologies. To make this statement precise would require a deeper
analysis of continuous relations in the context of convergent
topologies.\footnote{\ie~one wants to prove that $\odot$ is a cartesian
product in the category of ``localized interaction structures'' (see
definition~\ref{defn:localized} \vpageref{defn:localized}) with ``convergent
and total'' general (see~\cite{BPIV}) simulations.}

\subsubsection{Extending the execution formula}
\label{sec:extending}

The definition of basic topology places few constraints on the $\A$ and $\J$
operators.  Compatibility is a very weak requirement. On the other hand,
the~$\A_w$ and~$\J_w$ generated from an interaction structure $w$ have a lot
in common. In particular, \emph{classically}, $\A_w$ and $\J_w$ are dual:
$\comp\SEQ\A_w=\J_w\SEQ\comp$; and the positivity axiom is classically always
true!  It is thus natural to ask whether we can extend our interpretation to
take into account more basic topologies.  It is possible if we use different
interaction structures to generate the~$\A$ and the~$\J$:

\begin{prop}
  Suppose that $R$ is a simulation of $<S_h,w_h>$ by $<S_l,w_l>$. Then
\begin{enumerate}
  \item $\angelU{R}·\J_{l}·\demonU{\CONV{R}}$ is an interior
  operators on $\Pow(S_h)$;
\item $\A_{h}$ is compatible with $\angelU{R}·\J_{l}·
  \demonU{\CONV{R}}$.
\end{enumerate}
\ie~$\big(S_h\,,\,\A_h\,,\,\angelU{R}·\J_{l}·\demonU{\CONV{R}}\big)$ is a basic topology.
\end{prop}

\begin{proof}
First point: $\angelU{R}·\J_{l}·\demonU{\CONV{R}}$ is an interior operator:
\begin{itemize}
  \item $\angelU{R}·\J_{l}·\demonU{\CONV{R}}(U) \sub \angelU{R}·\demonU{\CONV{R}}(U) \sub U$

\item we have:

  \smallskip
  \begin{algproof}
    $\angelU{R}·\J_{l}·\demonU{\CONV{R}}(U)\sub V$

  \step{=>}{}

     $\demonU{\CONV{R}}·\angelU{R}·\J_{l}·\demonU{\CONV{R}}(U)\sub \demonU{\CONV{R}}(V)$

   \step{=>}{since $U\sub\demonU{\CONV{R}}·\angelU{R}(U)$}

     $\J_{l}·\demonU{\CONV{R}}(U)\sub \demonU{\CONV{R}}(V)$

   \step{=>}{$\J_l$ is an interior operator}

     $\J_{l}·\demonU{\CONV{R}}(U)\sub \J_l·\demonU{\CONV{R}}(V)$

   \step{=>}{}

     $\angelU{R}·\J_{l}·\demonU{\CONV{R}}(U)\sub\angelU{R}·\J_l·\demonU{\CONV{R}}(V)$
  \end{algproof}
\end{itemize}
This completes the proof that $\CONV{R}·\J_{l}·\demonU{\CONV{R}}$ is an
interior operator.

\medbreak
Second point: $\angelU{R}·\J_{l}·\demonU{\CONV{R}}$ is compatible with $\A_h$.
\\
Let $s_h|=\A_h(U)$ and $s_h|= \angelU{R}·\J_{l}·\demonU{\CONV{R}}(V)$, \ie~we
have an $s_l'$ \st~$(s_h,s_l')|=R$ and $s_l'|=
\J_{l}·\demonU{\CONV{R}}(V)$. In particular, $s_l'|= R\big(\A_h(U)\big)$ and
so $s_l'|= \A_l\big(R(U)\big)$ by lemma~\ref{lem:genSimChar}.
\\
We can apply the execution formula in $w_l$ to obtain a final state $s_l''|=
\angelU{R}(U)$ \st~$s_l''|= \J_l·\demonU{\CONV{R}}(V)$, \ie~there is an
$s_h'|= U$ \st~$(s_h',s_l'')|=R$, which implies that $s_h'|=
\angelU{R}·\J_{l}·\demonU{\CONV{R}}(V)$.

\end{proof}

An interactive reading is that for interaction to take place, the Angel and
Demon do not need to use exactly same dialect. If the Angel uses $w_h$ and the
Demon uses $w_l$, the Demon needs to interpret actions in $w_h$ in terms of
actions in $w_l$, and the Angel needs to interpret reactions in $w_l$ in terms
of reactions in $w_h$, \ie~we need to have a simulation from $w_h$ to $w_l$.
Note that because of the respective roles of the Angel and Demon, one never
needs to translate actions from the Demon or reaction from the Angel.

\smallbreak
In~\cite{complInducTop}, Silvio Valentini investigates the problem of
``completeness'' of inductively generated topologies.  It might be interesting
to investigate the operation described above in this context.

\subsection{Localization and distributivity}
\label{sec:localdist}

The basic topology obtained from an interaction structure is not in general
distributive. One way to obtain distributivity is to add a condition of
\emph{convergence} (page~\pageref{convergence}).  In~\cite{induc_top}, the
authors introduce the notion of ``localized'' axiom set which gives rise to
convergent basic topologies, \ie~formal topologies.

If $w=<A,D,n>$ is an interaction structure, a preorder $\leq$ on $S$ is said to
be localized if the following holds:
\[
  s' \leq s\ ,\ a\in A(s)
  \quad => \quad
  s' |= w\Bigg( \bigcup_{d\in D(s,a)} \SING{s[a/d]}  \BinDown
  \SING{s'}\Bigg)
\]
This implies in particular that $\geq$ is a linear simulation.

\medbreak
Suppose that $\leq$ is localized; if we extend the generating rules with
\begin{center}
  \infer{s'|= U & s\leq s'}{s|=\A(U)}{$\leq$-compat}. \label{compat-rule}
\end{center}
then the resulting lattice is distributive: this is one of the results
of~\cite{induc_top}. (The rules were slightly more complex because they had
to consider the positivity predicate and the positivity axiom).
Note that since $\leq$ is a preorder ---and as such, reflexive--- this rule is
a generalization of the reflexivity rule.

\smallbreak
The preorder is intended to represent \textit{a priori} the notion of
inclusion between basic opens. The smallest interesting preorder to consider
is the following: ``$s \leq s'$ iff $s|=\A\{s'\}$''. This preorder is
the saturation of the identity
and it appears implicitly in the definition of convergence.

\medbreak
The rest of this section is devoted to an analysis of the notion of
localization in the context of interaction structures, together with a
tentative computational interpretation. It culminates with an interpretation
of the notion of formal points in terms of server programs.

\subsubsection{Interaction structure with self-simulation}

The first step is to add a preorder on states, and to require it to be
well-behaved with respect to its parent interaction structure.

\begin{defn}
  An interaction structure with \newterm{self-simulation} on $S$ is a pair $<w,R>$
  where
  \begin{itemize}
    \item $w$ is an interaction structure on $S$;
    \item $R$ is a general simulation from $w$ to itself.
  \end{itemize}
\end{defn}

\begin{lem}
If $R$ is a simulation from $w$ to itself, then so is the reflexive
transitive closure of $R$.
\end{lem}
\begin{proof}
  This is a direct consequence of the following facts: identities are
  simulations, simulations compose (proposition~\ref{prop:simCategory}) and
  simulations are closed under unions (proposition~\ref{prop:simClosedUnion}).

\end{proof}
As a result, without loss of generality we can assume the self-simulation to be a preorder, and we
call it ``$\geq$'', with converse $\leq$. 
The meaning of ``$s\leq s'$'' is thus ``$s$ simulates $s'$ in
$w$''. 
We write $\downclosure{\{s\}}$ for the segment $(s \geq)$ (or $(\leq s)$) below
$s \in S$, and $\downclosure{U}$ for the downclosure 
$\angelU{\leq}(U)$ 
of $U : \Pow(S)$. 

\noindent
We have:
\begin{lem} \label{lem:downClosureCover}
  $s <| V$ implies $\downclosure{\{s\}} <| \downclosure{V}$.
\end{lem}
\begin{proof}
This is just an application of proposition~\ref{prop:linSimChar}.
\end{proof}

\smallbreak
\noindent
Two extreme examples of such self simulations are:
\begin{itemize}
  \item the empty relation, or the identity (its reflexive/transitive
    closure).  This is isomorphic to the case of normal interaction structures.

  \item $R == (\J_w(S)×S) \cup (S×\A_w(\emptyset))$. The intuition is
    that $(s_D,s_A)|=R$ iff the Demon can avoid deadlocks from $s_D$ or the
    Angel can deadlock the Demon from $s_A$.\footnote{A Demon deadlock is a
    pair $(s,a)$ such that $D(s,a)=\emptyset$.} Classically, this can be shown
    to be the biggest simulation (\ie~it is the union of all simulations) on
    an interaction structure and \textit{a~fortiori}, we have $R = \RTC{R} =
    \Sat{\RTC{R}}$.
\end{itemize}

\begin{remark}
  The second example can be seen as a constructive contrapositive of the
  following fact:
  \begin{lem}
    Let $\geq$ be a self-simulation on an interaction structure $<A,D,n>$ on
    $S$; suppose that $s\leq s'$ ($s$ simulates $s'$); we have:
    \begin{itemize}
      \item if the Demon can avoid deadlocks from $s'$ then he can also avoid
        deadlocks from~$s$ (\ie~$s' |>< S => s|>< S$);
      \item if the Angel can drive the Demon into a deadlock from $s'$, then
        she can do it from~$s$ (\ie~$s' <| \emptyset => s<| \emptyset$).
    \end{itemize}
  Classically, the two points are equivalent.
  \end{lem}
  The second simulation is (classically) equivalent to the one \emph{defined}
  from those properties (\ie~$(s,s')|= R $ iff $s'<|\emptyset =>
  s<|\emptyset$ iff $s'|><S => s|><S$).
\end{remark}


\subsubsection{Interaction structures and localization}

We now investigate the result of strengthening the condition to get full
localization.

\begin{defn} \label{defn:localized}
  Let $<w,\geq>$ be an interaction structure with self simulation on $S$;
  we say that $<w,\geq>$ is localized if the following holds:
  \[
    s_1 \leq s_2\ ,\ a_2\in A(s_2)
    \quad => \quad
    s_1   <|_w  \bigcup_{d_2\in D(s_2,a_2)} \SING{s_2[a_2/d_2]}  \BinDown
    \SING{s_1}
    \ \hbox{.}
  \]
\end{defn}
\noindent
This condition is slightly more general than the one from~\cite{induc_top} in
the sense that it considers general simulations rather than linear ones.  Note
also that in contrast to the notion of convergence from
definition~\ref{def:convergence}, this definition doesn't require equality:
$\downclosure{\SING{s}}$ is defined in terms of $\leq$.

\smallbreak
First, we need a lemma.

\begin{lem}
  Suppose $<w,\geq>$ is localized on $w$; then we have:
  \[
    s_1 \leq s_2\ ,\ a_2'\in \RTC{A}(s_2)
    \quad => \quad
    s_1   <|_w  \bigcup_{d_2'\in \RTC{D}(s_2,a_2')} \SING{s_2[a_2'/d_2']}  \BinDown
    \SING{s_1}
    \ \hbox{.}
  \]
\end{lem}
This means that the additional condition is well behaved with respect to the
RTC operation. (This is analogous to
point~\textit{\ref{lem:genSimChar:pt:RTC}} of lemma~\ref{lem:genSimChar}; and
indeed, the proof is very similar.)

\begin{proof}
Suppose that $s_1\leq s_2$ and let $a'_2\in \RTC{A}(s_2)$. We work by induction
on the structure of $a'_2$.

\begin{itemize}
\item if $a'_2 = \Exit$, this is trivial.

\item if $a'_2 = \Call(a_2,k_2)$: by localization, we know that
  \[
    s_1 <| \bigcup_{d_2} \SING{s_2[a_2/d_2]} \BinDown
                         \SING{s_1}
    \ \hbox{.}
  \]
  Let $s'_1|= \bigcup_{d_2} \SING{s_2[a_2/d_2]} \BinDown \SING{s_1}$, in
  particular $s'_1\leq s_2[a_2/d_2]$ for some $d_2\in D_2(s_2,a_2)$. We can
  apply the induction hypothesis for $s'_1\leq s_2[a_2/d_2]$ and $k_2(d_2)$ to
  obtain:
  \[
    s'_1
    \quad <|
    \bigcup_{d'_2\in \RTC{D}(s_2[a_2/d_2],k_2(d_2))}
      \SING{s_2[a_2/d_2]\big[k_2(d_2)/d_2'\big]}
      \BinDown
      \{s'_1\}
    \ \hbox{.}
  \]
  We have $\bigcup_{d'_2\in \RTC{D}(s_2[a_2/d_2],k_2(d_2))} \sub
  \bigcup_{d'_2\in \RTC{D}(s_2,a_2')}$ and
  $\downclosure{\SING{s_1'}}\sub\downclosure{\SING{s_1}}$ (because $s_1'\leq
  s_1$), which implies that the right hand side is thus included in
  $\bigcup_{d'_2\in \RTC{D}(s_2,a_2')} \SING{s_2[a_2'/d_2']} \BinDown
  \SING{s_1}$. By monotonicity, we get
  \[
    \bigcup_{d_2} \SING{s_2[a_2/d_2]} \BinDown \SING{s_1}
    \quad  <|
    \bigcup_{d'_2\in \RTC{D}(s_2,a_2')}\SING{s_2[a_2'/d_2']} \BinDown
                                       \SING{s_1}
    \ \hbox{.}
  \]
  We get the result by transitivity.
\end{itemize}
\end{proof}

\begin{lem}\label{lem:localized}
  If $<w,\geq>$ is a localized interaction structure, then $s <| U$
  implies $s <| U\BinDown\SING{s}$.
\end{lem}
\begin{proof}
  Let $s <| U$, \ie~there is a $a'$ in $\RTC{A}(s)$ s.t. $\bigcup_{d'\in\RTC{D}(s,a')}\SING{s[a'/d']}\sub U$.
  Since $\leq$ is reflexive, and by the previous lemma, we know that
  \[
    s <|
    \bigcup_{d'} \SING{s[a'/d']} \BinDown \SING{s}
    \ \hbox{.}
  \]
  The RHS is obviously included in $U\BinDown\SING{s}$,
  so we get the result by monotonicity.

\end{proof}

\begin{cor} \label{cor:coverDownclosure}
  If $<w,\geq>$ is a localized interaction structure, then 
  $U <| V$ implies $U <| U \BinDown V$.
\end{cor}

We will now check that convergence is satisfied for such a $(S,\A_{w},\leq)$.
\begin{prop} \label{prop:localizedConvergent}
  If $<w,\geq>$ is a localized interaction structure, then 
$s <| U$ and $s <| V$ jointly imply $s <| U\BinDown V$.
\end{prop}
\begin{proof}
  By lemma~\ref{lem:localized}, we know that $s <| U\BinDown\SING{s}$.  By
  lemma~\ref{lem:downClosureCover} we know that $\downclosure{\SING{s}} <|
  \downclosure V$, which implies $U \BinDown\SING{s} <| \downclosure V$; which
  (by corollary~\ref{cor:coverDownclosure}) implies $\downclosure {(U
  \BinDown\SING{s})} <| V \BinDown (U\BinDown\SING{s})$.

  Since we have $\downclosure{(U \BinDown V)}=U \BinDown V$, we can deduce
  that
  \[
    s
    \quad <| \quad
    U \BinDown\SING{s}
    \quad <| \quad
    U \BinDown V \BinDown \SING{s}
    \quad\sub\quad
    U \BinDown V
    \ \hbox{.}
  \]
\end{proof}

However, strictly speaking, the proof of lemma~\ref{lem:convDistr} doesn't
apply to the preorder context. Instead, we have to match the operator
generated by adding the $\leq$-compat rule (\vpageref{compat-rule}), and
define:
\begin{defn}  \label{defn:convISClosure}
  if $<w,\leq>$ is an interaction structure with self simulation, we write
  $\A_{w,\leq}$ for the predicate transformer
  $U |-> \A_{w}(\downclosure U)$, where $\downclosure U$ is the down-closure of $U$,
  \ie~$\downclosure{U} == \{s\ |\ (\exists
  s'|= U)\, s\leq s'\}$.
\end{defn}
The intuition is quite straightforward: if $s'$ can simulate a state $s$
($s'\leq s$) in~$U$, then $s'$ is ``virtually'' in $U$ as well. This is way to
say that our notion of simulation is semantically a real simulation.

\begin{cor}
  If $w$ is an interaction structure and $\leq$ a localized preorder on $S$;
  then the collection of open sets (\ie~the collection of $U$
  \st~$U=\A_{w,\leq}(U)$) is distributive.
\end{cor}

\subsubsection{Computational interpretation}

In the last section we merely transposed the definitions from \cite{induc_top}
to the context of interaction structures.  It is not however obvious how to
make computational sense of these definitions.   We now present an analysis of
the localization in computational terms.  The key idea is that to interpret
localization, one needs to adopt the perspective of the Demon (server, $|><$
operator), rather than that of the Angel (client, $<|$ operator).

\medbreak
For example, the computational content of lemma~\ref{lem:localized}
is that it is possible for the Angel to conduct interaction in such
a way that the behavior of the starting state can always be recovered by
simulation. The Angel takes care that she can at any point change her mind, and
abandon the current computation.  An example of a command which one would
\emph{not} have in such a system is ``\texttt{Reset}'', a command to brings
the whole system back to factory settings. This would lose all information
about previous interactions, which is not possible in a localized structure.

Thus, localization is a strong condition on interaction structures,
requiring them to be exceptionally well behaved. 

\begin{remark}
  Note that the notion of ``localization'' for a game has little to do with 
  the notion of backtracking present in \cite{coquandGames}, where a game-theoretical
  interpretation of classical logic is presented.

  That the Angel is allowed to backtrack means that the Angel can ``go back in
  the past''. If the game is localized, then the Angel does not return to a
  previous state, but plays in the current state ``as if it were'' the previous
  state. In particular, the Angel retains the right to make moves in
  the current state.
\end{remark}

\noindent
There is a problem with interpreting the proof of
proposition~\ref{prop:localizedConvergent}: a non-canonical choice was made.
In the proof, we decided to first execute the client corresponding to $s <|
U$, and then the client corresponding to $s <| V$ on top of it. The opposite
works just as well:

\[
  s
  \quad <| \quad
  V \BinDown \SING{s}
  \quad <| \quad
  V \BinDown U \BinDown \SING{s}
  \quad\sub\quad
  V \BinDown U
  \ \hbox{.}
\]
The two different witnesses for $s <| U \BinDown V$ may be quite different in
terms of execution! Here is what happens in graphical terms:
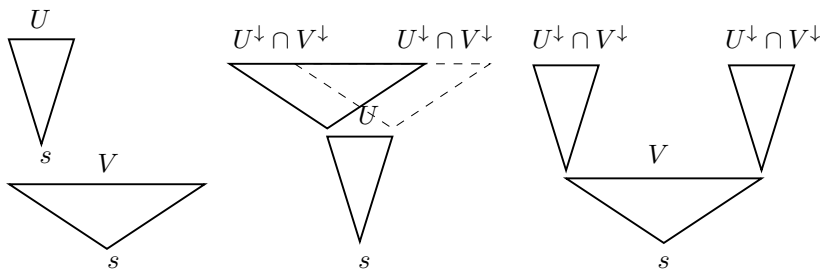
\begin{center}
  \begin{picture}(0,0)%
\includegraphics{compose_localize.pstex}%
\end{picture}%
\setlength{\unitlength}{3158sp}%
\begingroup\makeatletter\ifx\SetFigFont\undefined%
\gdef\SetFigFont#1#2#3#4#5{%
  \reset@font\fontsize{#1}{#2pt}%
  \fontfamily{#3}\fontseries{#4}\fontshape{#5}%
  \selectfont}%
\fi\endgroup%
\begin{picture}(6143,2073)(1542,-4240)
\put(1801,-3361){\makebox(0,0)[lb]{\smash{\SetFigFont{10}{12.0}{\familydefault}{\mddefault}{\updefault}{\color[rgb]{0,0,0}$s$}%
}}}
\put(1726,-2311){\makebox(0,0)[lb]{\smash{\SetFigFont{10}{12.0}{\familydefault}{\mddefault}{\updefault}{\color[rgb]{0,0,0}$U$}%
}}}
\put(6526,-3361){\makebox(0,0)[lb]{\smash{\SetFigFont{10}{12.0}{\familydefault}{\mddefault}{\updefault}{\color[rgb]{0,0,0}$V$}%
}}}
\put(6601,-4186){\makebox(0,0)[lb]{\smash{\SetFigFont{10}{12.0}{\familydefault}{\mddefault}{\updefault}{\color[rgb]{0,0,0}$s$}%
}}}
\put(4276,-4186){\makebox(0,0)[lb]{\smash{\SetFigFont{10}{12.0}{\familydefault}{\mddefault}{\updefault}{\color[rgb]{0,0,0}$s$}%
}}}
\put(4276,-3061){\makebox(0,0)[lb]{\smash{\SetFigFont{10}{12.0}{\familydefault}{\mddefault}{\updefault}{\color[rgb]{0,0,0}$U$}%
}}}
\put(2251,-3436){\makebox(0,0)[lb]{\smash{\SetFigFont{10}{12.0}{\familydefault}{\mddefault}{\updefault}{\color[rgb]{0,0,0}$V$}%
}}}
\put(2326,-4186){\makebox(0,0)[lb]{\smash{\SetFigFont{10}{12.0}{\familydefault}{\mddefault}{\updefault}{\color[rgb]{0,0,0}$s$}%
}}}
\put(7126,-2461){\makebox(0,0)[lb]{\smash{\SetFigFont{10}{12.0}{\familydefault}{\mddefault}{\updefault}{\color[rgb]{0,0,0}$\downclosure U \cap \downclosure V$}%
}}}
\put(5626,-2461){\makebox(0,0)[lb]{\smash{\SetFigFont{10}{12.0}{\familydefault}{\mddefault}{\updefault}{\color[rgb]{0,0,0}$\downclosure U \cap \downclosure V$}%
}}}
\put(4576,-2461){\makebox(0,0)[lb]{\smash{\SetFigFont{10}{12.0}{\familydefault}{\mddefault}{\updefault}{\color[rgb]{0,0,0}$\downclosure U \cap \downclosure V$}%
}}}
\put(3301,-2461){\makebox(0,0)[lb]{\smash{\SetFigFont{10}{12.0}{\familydefault}{\mddefault}{\updefault}{\color[rgb]{0,0,0}$\downclosure U \cap \downclosure V$}%
}}}
\end{picture}

\end{center}
On the left are the two client programs witnessing $s <| U$ and $s <| V$; and
on the right, the two different programs witnessing $s <| U\BinDown V$.

Even worse, when the programs corresponding to $s <| U$ and $s <| V$ are
non-trivial, we could interleave the programs before reaching $U \BinDown V$!

\bigbreak
To give computational sense to the notion of localization, consider a server
interacting with clients. We allow ourselves a degree of anthropomorphism, by
referring to what these parties ``believe''.

Think of the self-simulation $\geq$ as a relation between ``visible'' or
``virtual'' states for the client(s) and ``internal'' server states. Because
this is a (general) simulation, it is guaranteed that we can conduct
interaction in the following way:
\begin{enumerate}
  \item if $s'\leq s$, \ie~the Angel believes the Demon is in a state $s$, but
    internally, the Demon is really in a state $s'$ that simulates $s$;

  \item the Angel sends a request $a \in A(s)$;

  \item the Demon does the following:
    \begin{enumerate}
      \item translates the $a\in A(s)$ into a $a'\in\RTC{A}(s')$ (by
        simulation),
      \item responds to $a'$ with a $d'\in \RTC{D}(s',a')$ (because it is a
        server program),
      \item and translate this answer $d'$ into a $d\in D(s,a)$ (by
        simulation);
    \end{enumerate}

  \item The Angel receives the answer $d \in D(s,a)$;

  \item the Angel now believes the new state is $n(s,a,d)$ while
  internally, the Demon is really in state $\RTC{n}(s',a',d')\leq
  n(s,a,d)$ that simulates $n(s,a,d)$;
\end{enumerate}
In particular, after the last point, the Angel can continue interaction.

\smallbreak
Localization can then be seen as the following requirement: suppose the server
is internally in a state $s$ and that there are two clients who respectively
believe it is in state $s_1$ and $s_2$. The two clients can send their
requests and the server respond to them (as above) in any order.  Suppose the
server first responds the first client.  Then at point (a) in the analysis of
client-server interaction above, the server can chose some $a'$ which is
constrained to bring about a state $s'[a'/d']\leq s_1[a_1/d_1]$ (like above)
\emph{and} $s'[a'/d']\leq s$ (by localization).  The first condition allows
the first client to continue interaction, while the second point
(localization) guarantees that the server can also answer requests to the
second client (because $s'[a'/d']\leq s \leq s_2$)...

In other words, that an interaction structure is equipped with a localized
self-simulation means that we can construct ``concurrent   virtual servers''
with which several clients can interact independently.

\medbreak
One way to localize any interaction structure on $S$ is the following:
define $\L(w)$ on $\mathit{Fin}(S)$\footnote{where $\mathit{Fin}(S)$ is the
collection of finite subsets of $S$. Finite subsets are represented by
families indexed by a finite set (\ie~an integer).} as
\be
  \ISA{\L(w)}(\FAM{s_i}{i \in I})   &==&  \SI{i \in I }\ISA{w}(s_i)\\
  \ISD{\L(w)}(\FAM{s_i}{i \in I},<i,a>)   &==& \ISD{w}(s_i,a)\\
  \ISn{\L(w)}(\FAM{s_i}{i \in I},<i,a>,d) &==& \FAM{s_i}{i \in I} \cup \SING{s[a/d]}
\ee
with reverse inclusion as simulation order.
(To define the inclusion order between families of states of course requires
there to be an equality relation between states.) This interaction structure
$<\L(w),\geq>$ is automatically localized.

The idea is simply that the Demon keeps a log of all the previous states
visited during interaction, so that he can use any ``past'' state as the
current one.

\begin{remark}
  To get a situation which is even closer to ``real life'', one can define a
  simulation $R : w -> \L(w)$ with $(s,l) |= R$ iff ``$s|= l$'' and use the
  extension from section~\ref{sec:extending} to interpret interaction.

  The idea is that the Demon advertizes a service specified by the interaction
  structure $w$; but internally implements $\L(w)$ in order to deal with
  concurent requests. The clients are only supposed to use interface $w$.
\end{remark}

\paragraph{Points.}

Now that the notion of localized interaction structure (aka formal topology)
has a computational interpretation, we can look at the notion of formal point.
Recall that a formal point in a formal topology on $S$ is a subset $\alpha
\sub S$ such that (see~\cite{BPIV})
\begin{itemize}
  \item $\alpha$ is closed;
  \item $\alpha$ is non-empty ($\alpha\olp\alpha$);
  \item $\alpha$ is convergent, \ie~$s|=\alpha$ and $s'|=\alpha$ imply that
    $s\BinDown s' \olp \alpha$.
\end{itemize}

We have already described the computational interpretation of a closed subset
in section~\ref{sec:clientServer}: the closed subset $\J(V)$ is a
specification for a server program that can maintain $V$.\footnote{It is
straightforward to extend the $\J$ operator to the case of interaction
structures with self simulations: $\J_{w,\leq}(V) = \J_w(\upclosure{V})$.}
That it is non-empty means that we actually have a proof that $s|><\alpha$ for
some $s$, \ie~that we have a server program maintaining $V$ (from some
specific state $s$).

\smallbreak
\noindent
Thus, a point is nothing more than a specification for a server program that
satisfies
\begin{itemize}
  \item if $s_1 |= \alpha$ (a client may connect in state $s_1$)

  \item and if $s_2 |= \alpha$ (a client may connect in state $s_2$)

  \item then there is a(n internal) state $s$ that simulates both $s_1$ and
    $s_2$ (since $s|= s_1\BinDown s_2$) such that $s|=\alpha$. In other words,
    the server can find an internal state which will allow it to respond to both
    $s_1$ and $s_2$.
\end{itemize}

Formal points are thus ``coherent'' server program specifications in the sense
that they can satisfy any finite number of ``unrelated'' concurrent clients.

\paragraph{Continuous maps.}

A relation $R$ between two localized interaction structures $w_h$ and $w_l$ is
called a continuous map if we have (see~\cite{BPIV}):
\begin{itemize}
  \item $R$ is a general simulation from $w_h$ to $w_l$;
  \item $R$ is total: $S_l <|_{w_l,\leq} R(S_h)$;
  \item $R(s_1)\BinDown R(s_2) <|_{w_l,\leq} R(s_1\BinDown s_2)$.
\end{itemize}
Similar interpretation can be devised for continuous maps as for formal
points, but the relevance of this interpretation in terms of actual
client/server programming is still unclear. We prefer to leave the matter open
for the time being.



\section{Conclusion, and questions raised}

We hope to have shown that much of \emph{basic} topology has a natural
interpretation in programming terms.  On reflection this is not surprising:
programming is essentially about ``how to get there from here'', and this is a
notion with a topological flavor.

Our study of interaction structures began with the intention of clarifying
monotone predicate transformers such as those which model specifications in
imperative programming.  We have defined a category in which the objects
represent command-response interfaces, and the morphisms represent program
components that implement one (higher-level) interface ``on top of'' other
(lower-level) interfaces.  The category coincides with the category of basic
spaces and continuous relations.  Closure and interior operators are related
to server programs and client programs, and continuous maps to simulations of
one server ``on top of'' another.  We have tentatively proposed a computational
interpretation of those notions of formal topology connected with convergence,
and particularly the notion of point. 


We would also like to find topological counterparts of fundamental
computational notions.  For example, safety properties are essentially the
same as closed sets; but what about fairness properties?  For another example we
have seen that the notion of forward data refinement in programming is
connected with the notion of continuity (at least at the level of basic
topology).  From the computer science literature, it is known that both
forward data refinement and backward data refinement are required 
for refinement of abstract data types (see for example
\cite{GardinerMorgan93}).  Similar completeness properties hold in approaches
to refinement based on functions and auxiliary variables rather than relations
(see for example the use of history and prophecy variables as in
\cite{abadi91existence}).  It seems interesting therefore to ask whether
backward simulation or the use of prophecy variables has a topological
interpretation.

Another line of work concerns the model of classical linear logic presented
in~\cite{PTlinear}.

Finally, one hopes that the use of \emph{dependent} theory type permits the
expression of interface specifications with full precision ---that is, going
beyond mere interface signatures.  This might serve as a foundation for
designing components in real programming languages.  Tools to aid design might
be built on this foundation.  However examples of interfaces and simulations
are needed both to ensure that our model properly captures important properties of
interfaces, and also to find ergonomically smooth ways of working with simulations.

\paragraph{Acknowledgments.}


Thierry Coquand drew the authors' attention to the Peterson and Synek tree
sets and explained their connection with inductively defined coverage systems
in formal topology.  He stressed the importance of simulation.  He organized a
meeting with Sambin.  Both authors are indebted to him.  Sambin's work on
formal topology has been a delightful source of plunder.  The second author
also wants to thank Giovanni Sambin for inviting him twice: once for the
Second Workshop on formal topology in Venice; and once on a personal basis in
Padova to share ideas. The collaboration of Anton Setzer has helped the first
author's understanding of coinductive notions
(\cite{setzerhancock:venice2003,michelbrinksetzer:statedependentmonads:2004},
and of coalgebraic programming with dependent types
(\cite{MR2002h:68027,hancocksetzer:pontedelima:2000}).



\bibliographystyle{plain}
\bibliography{bibliography}
\end{document}